\RequirePackage{fix-cm}
\documentclass[smallextended]{svjour3}       % onecolumn (second format)
\smartqed  % flush right qed marks, e.g. at end of proof
\usepackage{graphicx}
\usepackage{xspace}
\usepackage{booktabs} % For better looking tables
\usepackage{siunitx} % For centering text in columns
\usepackage{tabularx} % For tables that exceed width of the page
\usepackage{xspace}
\usepackage{multirow}
\usepackage{listings}
\usepackage{enumitem}
\usepackage{tikz}
\usepackage{colortbl}
\usepackage{url}
\usepackage{cite}
\usepackage{caption}% added <<<<<<<<<<<<<<
\usepackage[colorlinks]{hyperref}
\hypersetup{hidelinks}
\usepackage[colorinlistoftodos,prependcaption,textsize=tiny]{todonotes}
\usepackage{natbib}
\usepackage{tcolorbox}
\usepackage{pdfpages}

\usepackage{float} % To use the [H] specifier
\definecolor{codegreen}{rgb}{0,0.6,0}
\definecolor{codegray}{rgb}{0.5,0.5,0.5}
\definecolor{codepurple}{rgb}{0.58,0,0.82}
\definecolor{backcolour}{rgb}{0.95,0.95,0.92}

\lstdefinestyle{mystyle}{
	commentstyle=\color{codegreen},
	keywordstyle=\color{magenta},
	numberstyle=\tiny\color{codegray},
	stringstyle=\color{codepurple},
	basicstyle=\small,
	breakatwhitespace=false,         
	breaklines=true,                 
	captionpos=b,                    
	keepspaces=true,                 
	numbers=left,                    
	numbersep=5pt,                  
	showspaces=false,                
	showstringspaces=false,
	showtabs=false,                  
	tabsize=4,
}

\definecolor{codegreen}{rgb}{0,0.6,0}
\definecolor{codegray}{rgb}{0.5,0.5,0.5}
\definecolor{codepurple}{rgb}{0.58,0,0.82}
\definecolor{backcolour}{rgb}{0.95,0.95,0.92}
\definecolor{backgrey}{rgb}{0.96,0.96,0.96}

\lstdefinestyle{mystyle}{
	basicstyle=\ttfamily\fontsize{8}{8}\selectfont,
	breakatwhitespace=false,         
	breaklines=true,                 
	captionpos=b,                    
	keepspaces=true,                 
	numbersep=5pt,                  
	showspaces=false,                
	showstringspaces=false,
	showtabs=false,                  
	tabsize=4
}

\lstdefinestyle{pstyle}{
	commentstyle=\color{codegreen},
	keywordstyle=\color{magenta},
	numberstyle=\tiny\color{codegray},
	stringstyle=\color{red},
	basicstyle=\ttfamily\fontsize{8}{8}\selectfont,
	breakatwhitespace=false,         
	breaklines=true,                 
	captionpos=b,                    
	keepspaces=true,                 
	numbers=left,                    
	numbersep=5pt,                  
	showspaces=false,                
	showstringspaces=false,
	showtabs=false,                  
	tabsize=4
}

\newcommand{\CodeSymbol}[1]{\textcolor{magenta}{#1}}
\lstdefinestyle{promptstyle}{
	backgroundcolor=\color{backgrey},   
	commentstyle=\color{black},
	keywordstyle=\color{black},
	stringstyle=\color{black},
	basicstyle=\linespread{1.1}\ttfamily\fontsize{8}{8}\selectfont,
	breakatwhitespace=false,         
	breaklines=true,                 
	captionpos=b,                    
	keepspaces=true,                 
	numbers=none,                    
	numbersep=1pt,                  
	showspaces=false,                
	showstringspaces=false,
	showtabs=false,                  
	tabsize=2,
    frame=bt,
    literate={\{}{{\CodeSymbol{\{}}}1
    {\}}{{\CodeSymbol{\}}}}1
}

\usepackage{tikz}
\usetikzlibrary{shapes, shapes.geometric, arrows, positioning}

\colorlet{punct}{red!60!black}
\definecolor{background}{HTML}{EEEEEE}
\definecolor{delim}{RGB}{20,105,176}
\colorlet{numb}{magenta}

\lstdefinelanguage{json}{
	basicstyle=\ttfamily\fontsize{8}{8}\selectfont,
    numbers=left,
    stepnumber=1,
    numbersep=8pt,
    numberstyle=\tiny\color{codegray},
    showstringspaces=false,
    breaklines=true,
    tabsize=4,
    literate=
     *{0}{{{\color{numb}0}}}{1}
      {1}{{{\color{numb}1}}}{1}
      {2}{{{\color{numb}2}}}{1}
      {3}{{{\color{numb}3}}}{1}
      {4}{{{\color{numb}4}}}{1}
      {5}{{{\color{numb}5}}}{1}
      {6}{{{\color{numb}6}}}{1}
      {7}{{{\color{numb}7}}}{1}
      {8}{{{\color{numb}8}}}{1}
      {9}{{{\color{numb}9}}}{1}
      {:}{{{\color{punct}{:}}}}{1}
      {,}{{{\color{punct}{,}}}}{1}
      {\{}{{{\color{delim}{\{}}}}{1}
      {\}}{{{\color{delim}{\}}}}}{1}
      {[}{{{\color{delim}{[}}}}{1}
      {]}{{{\color{delim}{]}}}}{1},
}

\lstdefinelanguage{JavaScript}{
  keywords={typeof, new, true, false, catch, function, return, null, catch, switch, var, if, in, while, do, else, case, break},
  keywordstyle=\color{blue}\bfseries,
  ndkeywords={class, export, boolean, throw, implements, import, this},
  ndkeywordstyle=\color{darkgray}\bfseries,
  identifierstyle=\color{black},
  sensitive=false,
  comment=[l]{//},
  morecomment=[s]{/*}{*/},
  commentstyle=\color{purple}\ttfamily,
  stringstyle=\color{red}\ttfamily,
  morestring=[b]',
  morestring=[b]"
}

\newcommand*{\priority}[1]{%
	\pgfmathsetmacro\percentage{#1} % Extract the percentage
	\begin{tikzpicture}[scale=0.125]%
		\draw (0,0) circle (1);
		\pgfmathparse{\percentage > 80 ? 1 : 0}
		\ifnum\pgfmathresult=1
			\fill[fill opacity=0.5,fill=green!60!black] (0,0) -- (90:1) arc (90:90-\percentage*3.6:1) -- cycle; % Green for percentages above 80
		\else
			\pgfmathparse{\percentage > 60 ? 1 : 0}
			\ifnum\pgfmathresult=1
				\fill[fill opacity=0.5,fill=blue!80!black] (0,0) -- (90:1) arc (90:90-\percentage*3.6:1) -- cycle; % Yellow for percentages between 60 and 80
			\else
				\pgfmathparse{\percentage > 40 ? 1 : 0}
				\ifnum\pgfmathresult=1
					\fill[fill opacity=0.5,fill=orange!90!black] (0,0) -- (90:1) arc (90:90-\percentage*3.6:1) -- cycle; % Blue for percentages between 40 and 60
				\else
					\pgfmathparse{\percentage > 20 ? 1 : 0}
					\ifnum\pgfmathresult=1
						\fill[fill opacity=0.5,fill=magenta!80!black] (0,0) -- (90:1) arc (90:90-\percentage*3.6:1) -- cycle; % Orange for percentages between 20 and 40
					\else
						\fill[fill opacity=0.5,fill=red!90!black] (0,0) -- (90:1) arc (90:90-\percentage*3.6:1) -- cycle; % Red for percentages below 20
					\fi
				\fi
			\fi
		\fi
	\end{tikzpicture}%
}

\newcommand*{\circlecolor}[1]{%
	\begin{tikzpicture}[scale=0.125]%
		\draw (0,0) circle (1);
		\ifnum#1=1
		\fill[fill opacity=0.7,fill=blue!70!black] (0,0) circle (1);
		\fi
	\end{tikzpicture}%
}

\newcommand{\headergen}{\textsc{HeaderGen}\xspace}
\newcommand{\typeevalpy}{\textsc{TypeEvalPy}\xspace}
\newcommand{\pycg}{\textsc{PyCG}\xspace}
\newcommand{\swarmcg}{\textsc{SWARM-CG}\xspace}
\newcommand{\swarmjs}{\textsc{SWARM-JS}\xspace}

\begin{document}

\title{An Empirical Study of Large Language Models for Type and Call Graph Analysis in Python and JavaScript
\thanks{Funding for this study was provided by the Ministry of Culture and Science of the State of North Rhine-Westphalia under the SAIL project with the grand no NW21-059D.}
}
% \subtitle{Do you have a subtitle?\\ If so, write it here}

\titlerunning{An Empirical Study of LLMs for Type and Call Graph Analysis}        % if too long for running head

\author{Ashwin Prasad Shivarpatna Venkatesh \and
        Rose Sunil \and
        Samkutty Sabu \and
        Amir M. Mir \and
        Sofia Reis \and
        Eric Bodden}

\institute{Ashwin Prasad Shivarpatna Venkatesh \at
              Heinz Nixdorf Institut, Paderborn University, Paderborn, Germany \\
              \email{ashwin.prasad@upb.de}
           \and
           Rose Sunil \at
              Paderborn University, Paderborn, Germany \\
              \email{rose10@mail.uni-paderborn.de}
              \and
           Samkutty Sabu \at
              Paderborn University, Paderborn, Germany \\
              \email{samkutty@mail.uni-paderborn.de}
           \and
           Amir M. Mir \at
              Delft University of Technology, Delft, The Netherlands \\
              \email{s.a.m.mir@tudelft.nl}
           \and
           Sofia Reis \at
              IST, University of Lisbon \& INESC-ID, Lisbon, Portugal \\
              \email{sofia.o.reis@tecnico.ulisboa.pt}
           \and
           Eric Bodden \at
              Heinz Nixdorf Institut \& Fraunhofer IEM, Paderborn University, Paderborn, Germany \\
              \email{eric.bodden@upb.de}
}

\date{Received: date / Accepted: date}

%\includepdf[pages=-]{journalrebuttal-major-revision.pdf}
\maketitle

\begin{abstract}
Large Language Models (LLMs) are increasingly being explored for their potential in software engineering, particularly in static analysis tasks.
In this study, we investigate the potential of current LLMs to enhance call-graph analysis and type inference for Python and JavaScript programs.
We empirically evaluated 24 LLMs, including OpenAI's GPT series and open-source models like LLaMA and Mistral, using existing and newly developed benchmarks.
Specifically, we enhanced TypeEvalPy, a micro-benchmarking framework for type inference in Python, with auto-generation capabilities, expanding its scope from 860 to 77,268 type annotations for Python.
Additionally, we introduced SWARM-CG and SWARM-JS, comprehensive benchmarking suites for evaluating call-graph construction tools across multiple programming languages.

Our findings reveal a contrasting performance of LLMs in static analysis tasks.
For call-graph generation, traditional static analysis tools such as PyCG for Python and Jelly for JavaScript consistently outperform LLMs.
While advanced models like mistral-large-it-2407-123b and gpt-4o show promise, they still struggle with completeness and soundness in call-graph analysis across both languages.
In contrast, LLMs demonstrate a clear advantage in type inference for Python, surpassing traditional tools like HeaderGen and hybrid approaches such as HiTyper.
These results suggest that, while LLMs hold promise in type inference, their limitations in call-graph analysis highlight the need for further research.
Our study provides a foundation for integrating LLMs into static analysis workflows, offering insights into their strengths and current limitations.
\end{abstract}

\section{Introduction}
In recent years, the field of Software Engineering (SE) has witnessed a paradigm shift with the integration of Large Language Models (LLMs), bringing new capabilities and enhancements to traditional software development processes~\citep{rasnayaka_empirical_2024,huang_generative_2024, houLargeLanguageModels2023, fanLargeLanguageModels2023, zhangSurveyLargeLanguage2023, zhengUnderstandingLargeLanguage2023}.
LLMs, with their ability to understand and generate human-like text, are reshaping various SE tasks, such as code generation and bug detection.

Static analysis (SA), a foundational technique in SE, focuses on evaluating code without executing it, allowing developers to detect potential errors, maintain code quality, and identify security vulnerabilities early in the development lifecycle.
Historically, SA tools have faced challenges, such as high rates of false positives, difficulty of scaling to large codebases, and limited ability to handle ambiguous or incomplete code.
Models such as BERT~\citep{devlinBERTPretrainingDeep2019a}, T5~\citep{raffelExploringLimitsTransfer2023}, and GPT~\citep{radfordLanguageModelsAre} have demonstrated potential in automating complex SA tasks~\citep{zhangSurveyLargeLanguage2023}.

Recent works have shown how different SA tasks can benefit from LLMs, such as false-positives pruning~\citep{10.1145/3611643.3613078}, improved program behavior summarization~\citep{liHitchhikerGuideProgram2023}, type annotation ~\citep{seidel2023learning}, and general enhancements in precision and scalability of SA tasks~\citep{liHitchhikerGuideProgram2023, mohajer_effectiveness_2024}, both fundamental issues of SA.

\textbf{Goal:} This study positions itself at the intersection of SA and LLMs, examining the effectiveness of LLMs in SA within SE.
It aims to \textbf{evaluate the accuracy of LLMs in performing specific SA tasks in Python and JavaScript programs}, such as call-graph analysis and type inference.
We focus on Python and JavaScript as they are dynamically typed languages, making them inherently challenging for static analysis due to the absence of explicit type annotations and the high degree of runtime flexibility.
\textit{Call-graph analysis} helps in understanding the relationships and interactions between different components of a program, while \textit{type inference} aids in identifying potential type errors and improving code reliability.

\textbf{Methodology:} 
We performed a comprehensive analysis of the capabilities of $24$ different LLMs across different SA tasks, using data from micro-benchmarks and customized prompts for each task. This evaluation enables one to make direct comparisons with the existing capabilities of traditional approaches in SA.
To assess the performance of LLMs, we use the \pycg~\citep{salisPyCGPracticalCall2021c} and 
\headergen~\citep{venkateshEnhancingComprehensionNavigation2023a} micro-benchmarks for call-graph analysis in Python, and a newly created \textit{SWARM-JS} micro-benchmark for JavaScript.
For type inference, we use the \typeevalpy~\citep{venkateshTypeEvalPyMicrobenchmarkingFramework2023} micro-benchmark. 
The use of micro-benchmarks in evaluating the performance of LLMs in our study is grounded in the following key considerations:
\begin{itemize}
    \item Micro-benchmarks are designed to target specific aspects of the features under test and various characteristics of the programming language involved. 
    This helps highlight the models' strengths and weaknesses, allowing for a more nuanced understanding of their capabilities in SA tasks.
    \item Micro-benchmarks development involves rigorous manual inspection and adherence to scientific methods, ensuring reliability and accuracy in evaluation. 
Conversely, obtaining large-scale, real-world data that can serve as ground truth is often a challenging endeavor.
Where such data is available, it is susceptible to human errors, which can skew the results \citep{digraziaEvolutionTypeAnnotations2022}.
\end{itemize}

The insights from this study are intended to offer a preliminary understanding of the role of LLMs in SA for the call-graph analysis and type inference tasks, contributing to the Artificial Intelligence for Software Engineering (AI4SE) and Software Engineering for Artificial Intelligence (SE4AI) fields.  

\textbf{Results:}
The results of our study show that static analysis tools like PyCG and Jelly~\citep{laursen_reducing_2024} significantly outperform LLMs in call-graph generation for Python and JavaScript, respectively.
LLMs, while showing some promise, especially with models like mistral-large-it-2407-123b and gpt-4o, struggled with completeness and soundness in both Python and JavaScript.

Contrarily, in the case of type inference, LLMs demonstrated a clear advantage over traditional static analysis tools like \headergen and hybrid approaches such as HiTyper.
While OpenAI's gpt-4o initially performed best in the micro-benchmark, mistral-large-it-2407-123b surpassed it on the larger autogen-benchmark, indicating that some open-source models can outperform proprietary ones.

\textbf{Contributions:}
The primary contributions of this paper are as follows:
\begin{itemize}
    \item Performed an empirical evaluation of 24 LLMs across Python and JavaScript for call-graph inference.
    \item Conducted an empirical evaluation of 24 LLMs for type inference in Python.
    \item Enhanced \typeevalpy with auto-generation capabilities, expanding its benchmarking scope for type inference from 860 to 77,268 annotations.
    \item Introduced \swarmcg, a comprehensive benchmarking suite for evaluating call-graph construction tools across multiple programming languages, starting with Python and JavaScript, to enable cross-language comparisons and consistent analysis evaluations.
    \item Developed SWARM-JS, a call-graph micro-benchmark for JavaScript.
\end{itemize}

\textbf{Structure:} 
The structure of the paper is as follows: in Section~\ref{sec:background} we discuss the necessary background information, including tools and baselines used in the study.
In Section~\ref{sec:relatedwork} we discuss the related work.
The research questions are outlined in Section~\ref{sec:rq}.
In Section~\ref{sec:micro}, we describe the micro-benchmarks used for evaluation, while Section~\ref{sec:methodology} describes our methodology.
The results are presented in Section~\ref{sec:results} and subsequently discussed in Section~\ref{sec:discussion}. 
Section~\ref{sec:ttv} addresses the threats to validity.
Finally, the paper is concluded by outlining future research directions in Section~\ref{sec:conclusion}.

\textbf{Availability:}
\begin{itemize}
    \item \typeevalpy is published on GitHub as open-source software:
\url{https://github.com/secure-software-engineering/TypeEvalPy}
    \item \swarmcg is published on GitHub as open-source software: \url{https://github.com/secure-software-engineering/SWARM-CG}
    \item The raw outputs and analysis data are published on Zenodo at: \url{https://zenodo.org/records/15045642}
\end{itemize}

\section{Background}
\label{sec:background}

Program analysis techniques, such as \textit{type inference} and \textit{call-graph analysis}, are essential for static analysis tools that allows them to reason about program correctness before execution, especially in dynamically typed languages. 
Type inference determines the types of variables without explicit annotations, while call-graph analysis tracks function calls and their relationships within a program. 

\textbf{Type inference.}
Type inference is the process of deducing the types of variables based on available program information, such as function signatures and variable assignments.
In dynamically typed languages like Python, where variable types can change at runtime, type inference helps predict potential type mismatches and enforce consistency in operations.

A static analyzer with type inference capabilities examines assignments, function calls, and operations to deduce the type of each variable at different points in the program. 
By performing this analysis, type inference can detect errors such as type mismatches before execution, preventing runtime failures.

\textbf{Call-graph Analysis.}
A call-graph is a representation of function call relationships in a program. 
Call-graph analysis helps in understanding control flow, tracking function dependencies, and identifying unreachable or unused code.
It enables optimizations, refactoring, and bug detection by revealing function relationships.

\textbf{Flow-insensitive vs. Flow-sensitive Analysis.}
Flow-insensitive and flow-sensitive analyses differ in how they account for the order of program execution. 
Flow-\textit{insensitive} analysis disregards the sequence of statements, treating all potential assignments to a variable as if they occur simultaneously.
This lack of ordering leads to an over-approximation of possible program states, potentially reducing the precision of the analysis.
In contrast, flow-\textit{sensitive} analysis explicitly considers the order in which statements are executed, allowing it to track variable values and types throughout the program.
As a result, flow-sensitive approaches often yield more precise analysis outcomes.

\subsection{Motivating Example}

In the following code, the \texttt{create\_str} function returns a string, the variable \texttt{func\_ref} is assigned with function references at lines 4 and 8, and \texttt{x} is assigned the value \texttt{result + 1} at lines 6 and 10. 

\begin{lstlisting}[language=Python, style=pstyle]
def create_str(x):
	return x.upper()

func_ref = create_str
result = func_ref("Hello!")
x = result + 1 # Type mismatch!

func_ref = len
result = func_ref("Hello!")
x = result + 1 # Works

x = eval("create_str('Hello!')")
x = x + 1 # Type mismatch!
\end{lstlisting}

\textbf{Type Inference.}
A static analyzer with type inference capabilities can resolve that the variable \texttt{result} at line 5 is a string, while the variable \texttt{result} at line 9 is an integer.
Using this, the static analyzer can raise a type error at line 6 even before executing it.
However, static analyzers struggle with dynamically evaluated expressions that obscure type information at analysis time, such as the \texttt{eval} function in line 12.
If a variable’s value is determined through eval, reflection, or user input, static analyzers cannot reliably infer its type before execution.
This further highlights the challenges of type inference in dynamically typed languages, where runtime behavior can introduce unexpected type inconsistencies.

\textbf{Callgraph.}
The complete call-graph for the snippet is as follows:

\textit{main $\rightarrow$ create\_str() $\rightarrow$ upper()}

\textit{main $\rightarrow$ len()}

A flow-sensitive analysis can further resolve exactly where these calls are made.
For instance, it can resolve that at line 5 the variable \texttt{func\_ref} points to the function \texttt{create\_str} while at line 9 \texttt{func\_ref} points to the function~\texttt{len}.

\subsection{Type Inference Tools}
In this subsection, we briefly describe the existing type inference approaches, namely, Type4Py, HiTyper, and \headergen.

\textbf{Type4Py.} 
Type4Py~\citep{mirType4PyPracticalDeep2022c} is a deep similarity learning-based type inference model for Python that addresses the limitations of previous ML type inference methods.
Unlike earlier models trained on potentially incorrect human-provided annotations, Type4Py uses a type-checked dataset, ensuring higher accuracy in predicting variable, argument, and return types.
It maps programs into type clusters in a high-dimensional space, leveraging a hierarchical neural network model to differentiate between similar and dissimilar types.
This approach allows Type4Py to handle an unlimited type vocabulary. 
It improves upon the state-of-the-art models Typilus~\citep{Typilus} and TypeWriter~\citep{Typewriter}, achieving a Mean Reciprocal Rank (MRR) of 77.1\%, an 8.1\%, and 16.7\% improvement over these models, respectively.

To aid developers in retrofitting type annotations, Type4Py incorporates identifiers, code context, and visible type hints as features for type prediction.
Its deep similarity learning methodology enables the model to learn from a wide range of data, making it particularly effective for real-world usage.
Type4Py is also deployed as a Visual Studio Code extension, offering machine learning-based type auto-completion for Python, thereby enhancing developer productivity by easing the process of adding type annotations to existing codebases.

\textbf{HiTyper.}
HiTyper~\citep{HiTyper} is a hybrid type inference approach that combines static type inference with deep learning (DL) to address the challenges of type inference in dynamic languages like Python.
It builds on the observation that static inference can accurately predict types with static constraints, while deep learning models are effective for cases with dynamic features but lack type correctness guarantees.
HiTyper introduces a type dependency graph (TDG) to encode type dependencies among variables in a function.
By leveraging TDGs, HiTyper integrates static inference rules and type rejection rules to filter incorrect neural predictions, conducting an iterative process of static inference and DL-based type prediction until the TDG is fully inferred.

HiTyper’s key advantage lies in its ability to combine the precision of static inference with the adaptability of learning-based predictions. 
By focusing on \textit{hot type slots}, variables at the beginning of data flow with dependencies, HiTyper invokes DL models only when static inference is insufficient. 
Its similarity-based type correction algorithm supplements DL predictions, particularly for user-defined and rare types, which are challenging for traditional DL models. 
The results show that HiTyper outperforms state-of-the-art DL models like Typilus and Type4Py, achieving 10-12\% improvements in overall type inference accuracy and significant gains in inferring rare and complex types.

\textbf{HeaderGen.}
\headergen~\citep{venkateshEnhancingComprehensionNavigation2023a} is a static analysis-driven tool that analyzes Python code in Jupyter Notebooks to create structure and enhance the comprehensibility of the notebook.
\headergen uses a flow-sensitive call-graph analysis technique to extract fully qualified function names of all invoked function calls in the program and use this information as context to add structural headers to notebooks.
To facilitate flow-sensitive call-graph construction, the underlying analysis first constructs an assignment graph that captures the relationship between program identifiers.
\headergen augments this assignment graph with type information during its fixed-point iterations to infer types of program identifiers. 
The evaluation of \headergen on micro-benchmarks shows a precision of 95.6\% and a recall of 95.3\%. 

\subsection{Call-graph Construction Tools}

In this subsection, we briefly describe the existing call-graph generation approaches, namely, PyCG, \headergen, TAJS, and Jelly.

\textbf{PyCG.}
PyCG~\citep{salisPyCGPracticalCall2021c} is a static call-graph construction technique for Python.
It works by computing assignment relations between program, such as functions, variables, classes, and modules, through an inter-procedural analysis.
PyCG is capable of handling complex Python features such as modules, generators, lambdas, and multiple inheritance.
PyCG is evaluated on both micro-benchmarks and real-world Python packages.
PyCG outperforms other tools in terms of precision and recall, achieving a precision of 99.2\% and a recall rate of around 69.9\%.

\textbf{HeaderGen.}
As previously discussed, \headergen uses a flow-sensitive analysis to extract all invoked function calls within a program.
\headergen extends the assignment graph of PyCG to include flow-sensitive information, thereby increasing the precision of the call-graph construction algorithm.

\textbf{TAJS - Type Analyzer for JavaScript.}
TAJS~\citep{10.1007/978-3-642-03237-0_17} is a static analysis tool designed for JavaScript that performs type inference and constructs call-graphs.
It fully supports ECMAScript 3rd edition and provides partial support for ECMAScript 5, including its standard library, HTML DOM, and browser APIs.
However, TAJS does not support features introduced in ECMAScript 6~\citep{ecma-262}, such as classes, arrow functions, and modules, which limits its effectiveness in analyzing modern JavaScript applications.

TAJS offers a command-line option to export these call-graphs as DOT files, which can be converted into JSON for further analysis or integration with other tools.

\textbf{Jelly.}
Jelly~\citep{laursen_reducing_2024} is a hybrid JavaScript call-graph analysis tool that combines static and dynamic analysis to improve accuracy. 
Jelly’s analysis consists of two main steps. 
First, a dynamic pre-analysis is conducted to gather runtime hints regarding variable values and object structures. 
The second step uses these hints in a SA phase, refining the constructed call-graph and improving soundness. 
This method allows Jelly to outperform traditional SA tools, particularly in handling modern ECMAScript features and multi-file JavaScript programs.
Evaluations show that Jelly outperforms tools like TAJS\citep{10.1007/978-3-642-03237-0_17} and ACG\citep{feldthaus_efficient_2013}, making it more accurate for real-world JavaScript analysis.

% \textbf{js\_callgraph.}
% \begin{itemize}
%     \item The js-callgraph, \cite{10.5555/2486788.2486887}  implements a field-based algorithm for constructing call-graphs for JavaScript.
%     \item It has support for ECMAScript 6th edition features like arrow functions, destructuring assignments, classes, enhanced object literals and rest/spread operator.
%     \item The tool utilizes a flow analysis that tracks only the flow of function values rather than the flow of all objects.
%     \item The tool can be run in two basic modes, pessimistic and optimistic which can be selected using the --strategy flag.
%     \item The two modes differ in the way inter-procedural flows are handled. 
%     \item In the NONE strategy of basic pessimistic approach inter-procedural flow is not tracked at all.
%     \item The ONESHOT strategy in pessimistic mode is the default strategy and is a slight refinement of basic pessimistic approach where inter-procedural flow is tracked only for one-shot closures that are invoked immediately. 
%     \item The DEMAND strategy in optimistic mode performs inter-procedural propagation along edges that lead to a call site, enhancing call-graph construction.
%     \item The FULL strategy of optimistic mode that performs full inter-procedural propagation is not implemented yet.
% \end{itemize}

\section{Related Work}
\label{sec:relatedwork}

This section reviews prior work at the intersection of Large Language Models and static analysis, highlighting existing approaches, their limitations, and how our study advances the state of the art.

\subsection{Traditional Static Analysis for Python and JavaScript}

\textbf{Static Analysis Basics:} Static analysis (SA) tools examine code without executing it to find bugs, type errors, or security issues. In Python and JavaScript, traditional static analyzers include linters and type checkers. In Python, tools like \textbf{PyLint} and \textbf{Flake8} catch style issues and simple bugs, while \textbf{MyPy} and Facebook’s \textbf{Pyre} perform static type checking using type hints. JavaScript relies on linters (e.g., \textbf{ESLint}) and optional type systems (Flow or TypeScript’s compiler) to detect errors. Academic tools also exist: e.g., \textbf{PyCG} constructs call-graphs for Python using a context- and flow-insensitive analysis~\citep{salisPyCGPracticalCall2021c}, and \textbf{Type4Py} uses deep learning to predict Python types for better static type inference~\citep{mirType4PyPracticalDeep2022c}. These tools improve code quality but face well-known challenges with dynamic language features. Python and JavaScript allow dynamic typing, runtime reflection, and polyglot patterns that make static reasoning difficult. As a result, static analyzers often miss issues or report false alarms due to incomplete information. For instance, static taint analyzers depend on manually provided specifications of library APIs (sources/sinks), which are often missing or outdated, leading to missed vulnerabilities~\citep{li2025irisllmassistedstaticanalysis}. They also tend to over-approximate program behaviors, yielding many false positives that developers must triage~\citep{li2025irisllmassistedstaticanalysis}. In short, traditional SA tools are powerful but limited by dynamic features and scalability issues, motivating the exploration of learning-based approaches to augment or replace them.

\noindent\textbf{Advances in Static Analysis (pre-LLM):} Before the recent LLM surge, researchers began injecting machine learning into static analysis. Early deep neural network models trained on code graphs or tokens could predict types or detect certain bugs, but they lacked broad language understanding. For example, \textbf{DeepTyper}~\citep{10.1145/3236024.3236051} and \textbf{Typilus}~\citep{Typilus} learned to suggest variable types in Python, and \textbf{HiTyper} combined static inference with a neural network to improve Python type predictions~\citep{HiTyper}. These approaches showed that learned models can complement rule-based analysis, but they were narrow in scope and struggled with long-range dependencies in code.

\subsection{LLMs Enter Static Analysis: Early Experiments}

The emergence of code LLMs, such as OpenAI Codex, GPT-3, GPT-4 and Code LLaMa, prompted researchers to ask how well these models can perform classic static analysis tasks and where they fall short. Initial studies found that \textbf{LLMs easily handle basic syntax and code summarization but struggle with deeper program analysis}~\citep{sun2023automatic}, e.g., pointer analysis or detailed code behavior reasoning. Another survey noted that even large code models failed to reliably perform multi-step reasoning needed for vulnerability detection, achieving only ~55\% accuracy on such tasks without assistance~\citep{steenhoek2025errmachinevulnerabilitydetection}. In other words, LLMs are not ready to replace a static analyzer for complex analysis, as they often \textbf{hallucinate} facts or miss subtle relationships, especially when whole-program context is required.

On the positive side, researchers observed that LLMs have strong general knowledge of programming and can interpret code more semantically than traditional tools. ~\cite{liHitchhikerGuideProgram2023} argued that LLMs can be integrated into program analysis pipelines (“LLift”) to compensate for static analysis blind spots (e.g., LLMs might naturally summarize what a code segment does, helping an analyzer decide if a warning is a true issue). ~\cite{10.1145/3611643.3613078} took a first step in this direction by empirically testing ChatGPT as an assistant to static analysis. Overall, early investigations converged on the view that \textbf{LLMs are promising but have clear limitations}: they can understand intent and context in code, yet need careful prompting or fine-tuning to handle precise static analysis tasks. This realization spurred a new wave of techniques combining traditional static analysis strengths with LLMs’ flexibility. In this study, we investigate whether LLMs can effectively perform type inference and call-graph construction.

\subsection{LLM-Augmented Type Inference and Call Graph Construction}

Recent work has explored the application of LLMs to traditional static analysis tasks such as \textbf{type inference} and \textbf{call-graph construction}, particularly in dynamic languages like Python and JavaScript.

\cite{venkatesh_emergence_2024} conducted a comprehensive evaluation of $26$ models, including GPT-3.5, GPT-4 and Code LLaMA, on Python benchmarks targeting two core static analysis tasks: \textbf{type inference} and \textbf{call-graph construction}.
They found that \textbf{LLMs greatly outperformed a traditional analyzer on type inference accuracy, but lagged on call-graph construction}. In fact, GPT-4 inferred types in Python with higher accuracy than static methods, thanks to its learned knowledge of APIs and naming conventions. However, the same model struggled to predict dynamic function call targets, missing many edges in call-graphs. This suggests that LLMs excel at tasks requiring knowledge of common coding patterns (e.g. inferring that \texttt{len()} returns an \texttt{int}) but have trouble with exhaustive program structure analysis. The authors note that fine-tuning LLMs specifically for call-graph tasks or integrating them with algorithmic analysis might be necessary to overcome these limitations. Similarly, \cite{seidel2023learning} (CodeTIDAL) focused on TypeScript and trained a Transformer to predict missing type annotations, effectively learning from code context to enhance dataflow analysis. These efforts show that \textbf{LLMs can learn type and flow rules} in practice, often outperforming purely static approaches on inferring types, but they may need augmentation (e.g. external reasoning steps) for full program understanding. \cite{venkateshEnhancingComprehensionNavigation2023a} built \textbf{HeaderGen} to improve Python notebook analysis by adding flow-sensitive and type-aware reasoning on top of PyCG. HeaderGen demonstrated that adding semantic analysis and basic inference layers can improve static call-graph accuracy in Jupyter notebooks, even before full LLM integration. This underscores the potential of enhancing static tools with \textbf{LLM-augmented hybrid approaches}. This study extends \cite{venkatesh_emergence_2024} analysis to the JavaScript language.

\subsection{Micro-Benchmark Suites for Python and JavaScript}

\textbf{Static Call Graph Analysis Benchmarks in Python:}~\cite{salisPyCGPracticalCall2021c} introduced one of the first modern micro-benchmark suites for Python call-graph analysis as part of the PyCG study. This suite contains minimal Python programs covering a wide range of language constructs, organized into different feature-focused categories (e.g., simple function calls, decorators, generators, etc). The PyCG benchmarks provided a standardized way to compare static analyzers on Python’s dynamic features (e.g. lambdas, closures, dynamic dispatch via lists/dicts of functions) in a controlled setting. A strength of this suite is its breadth of coverage across Python 3’s core features and the inclusion of expected outputs for each test, which improves reproducibility and fair comparison. However, by design it uses only small, self-contained programs, this micro-scale yields clarity but may not capture interactions present in large codebases (e.g. extensive module interplay or runtime reflection beyond basic imports). Subsequent work built on PyCG’s benchmark to improve coverage and address its limitations. ~\cite{huang_scalable_2024} extended the suite with more snippets in their Jarvis call-graph analysis (adding 23 new tests on top of PyCG’s original 112). These additions include flow-sensitive scenarios, alongside other cases written by experienced Python developers to cover features that PyCG’s suite lacked. \cite{venkateshEnhancingComprehensionNavigation2023a} similarly created a micro-benchmark for their HeaderGen tool by adopting PyCG’s full suite and augmenting it with new snippets focused on flow-sensitive call sites. Both Jarvis’s and HeaderGen’s benchmarks remain centered on Python, inheriting the original PyCG test design and ground truths. 

\noindent\textbf{JavaScript Call Graph Analysis Benchmarks:} The SunSpider benchmark (WebKit 2010), a collection of JS programs originally meant for performance testing, has been used to compare call-graph tools, although it did not provide official ground-truth call-graph. Researchers had to manually inspect whether the edges produced by a tool match the actual calls in the source, a tedious process that focuses on precision of found edges and neglects recall (missing edges). ~\cite{Antal2023IsJavaScript} followed this approach in a comparative study, manually validating tool outputs on SunSpider. Such ad hoc methods are labor-intensive and error-prone, and they struggled to exercise modern JavaScript features beyond the aging SunSpider suite. Notably, a static analyzer (TAJS) showed high precision on classic benchmarks but failed to handle many ES6+ features, leading to underperformance on contemporary code. 

The lack of structured and standardized call-graph benchmarks across diverse
programming languages poses several challenges in evaluating and comparing call-graph construction tools. This gap makes cross-language comparisons
difficult and unreliable, hindering consistent assessments of different analysis
techniques. This study enables call-graph construction tools evaluation across multiple programming languages (SWARM-CG).

\noindent\textbf{Python Type Inference Benchmarks:} Researchers either used large-scale corpora of real code with optional type hints (e.g. the ManyTypes4Py dataset and Type4Py) or relied on each tool’s own set of examples, making it difficult to compare results across studies. Recognizing this gap, ~\cite{venkateshTypeEvalPyMicrobenchmarkingFramework2023} proposed TypeEvalPy, a micro-benchmarking framework for Python type inference tools. Categories cover dynamic typing constructs (uses of Python’s dynamic features that affect types, e.g. changing a variable’s type or using reflection) and external library calls, which simulate inferring types when third-party code is involved. While TypeEvalPy greatly improved standardization in evaluating Python type inference, it initially had some limitations in scope. The 154 snippets were designed to be representative but necessarily cannot cover all possible Python idioms. To address this, TypeEvalPy was augmented with an auto-generation extension that massively scaled up the benchmark’s coverage~\citep{venkatesh_emergence_2024}. This study describes the methodology to expand the synthetic test cases with $~77k$ type annotations that increase the diversity of types and scenarios.

\section{Research Questions}
\label{sec:rq}

We focus on the following research questions to evaluate the effectiveness of LLMs using micro-benchmarks in static analysis tasks: 

\begin{itemize}	
	\item[\textit{\textbf{RQ1:}}] \textit{What is the accuracy of LLMs in performing callgraph analysis against micro-benchmarks?}
	\item[\textit{\textbf{RQ2:}}] \textit{What is the accuracy of LLMs in performing type inference against micro-benchmarks?}
\end{itemize}

\section{Micro-benchmarks}
\label{sec:micro}
In this study, we utilize a diverse set of benchmarks to evaluate the effectiveness and performance of call-graph generation and type inference tools.
To support these evaluations, we extended existing frameworks and developed new benchmarks as required to ensure comprehensive testing.

To answer \textit{RQ1}, we choose two benchmarks designed to evaluate callgraph analysis performance, \pycg~\citep{salisPyCGPracticalCall2021c} and \headergen~\citep{venkateshEnhancingComprehensionNavigation2023a}.
Furthermore, we evaluate the effectiveness of LLMs across programming languages by creating a new micro-benchmark for JavaScript.

To answer \textit{RQ2}, we choose the micro-benchmark from \typeevalpy~\citep{venkateshTypeEvalPyMicrobenchmarkingFramework2023}, a general framework for evaluating type inference tools in Python.
\typeevalpy contains a micro-benchmark with 154 code snippets and 860 type annotations as ground truth.
Additionally, we extend the \typeevalpy with auto-generation capability and synthetically scale the micro-benchmark to include a wide spectrum of types.
The \typeevalpy \textit{autogen-benchmark} now contains 7,121 test cases with 77,268 type annotations.

In the following sections, we outline the micro-benchmarks used for both call-graph generation and type inference.

\subsection{PyCG: Call-graph Micro-benchmark}
The PyCG micro-benchmark suite offers a standardized set of test cases for researchers to evaluate and compare call-graph generation techniques. 
It includes 112 unique and minimal micro-benchmarks, each designed to cover different features of the Python language.
These benchmarks are grouped into 16 categories, ranging from simple function calls to more complex constructs like inheritance schemes.

Each category comprises multiple tests, with each test providing: 
(1) the source code, 
(2) call-graph in JSON format, and 
(3) a brief description of the test case. 
The tests are structured to be easy to categorize and expand, with each focusing on a single execution path, without the use of conditionals or loops.
This design ensures that the generated call-graph accurately reflects the execution of the source code.

To ensure completeness and quality, the authors had two professional Python developers review the suite, providing feedback on feature coverage and overall quality. 
Based on their recommendations, the authors refactored and further enhanced the suite.

In this study, we additionally include 14 new test cases to the PyCG benchmark based on the benchmark used in Jarvis~\citep{huang_scalable_2024}.
These additions include 4 in \textit{args} category, 4 in \textit{assignments}, 5 in \textit{direct\_calls}, and 1 in \textit{imports}.

\subsection{HeaderGen: Flow-sensitive Call-sites Micro-benchmark}

The micro-benchmark in \headergen is created by adopting the PyCG micro-benchmark.
\headergen adds flow-sensitive call-graph information, i.e., line number information indicating where in the program the call is originating from.
Furthermore, since \headergen performs a flow-sensitive analysis, eight new test cases specifically targeting flow-sensitivity are added.

\subsection{SWARM-CG: Swiss Army Knife of Call Graph Benchmarks}
The lack of structured and standardized call-graph benchmarks across diverse programming languages poses several challenges in evaluating and comparing call-graph construction tools.
This gap makes cross-language comparisons difficult and unreliable, hindering consistent assessments of different analysis techniques.

To address this issue, we developed the \textit{Swiss Army Knife of Call Graph Benchmarks (\swarmcg)}, a benchmarking suite designed to provide a standardized platform for evaluating call-graph construction tools across multiple programming languages. 
The primary goal of \swarmcg is to create a unified environment that facilitates consistent comparisons and promotes further research in the field of call-graph analysis, especially in the current landscape, where ML models are being explored as alternatives to traditional static analysis.
ML models often lack the transparency and verifiability that static analysis provides. 
As researchers investigate these models in call-graph construction, having a standardized framework is essential for accurately comparing their effectiveness with established methods.
\swarmcg fulfils this need by offering a well-organized, comprehensive set of call-graph benchmarks with ground truth annotations for each code snippet, enabling reliable and consistent evaluations.

Furthermore, each tool that \swarmcg supports is dockerized to make the evaluation process straightforward.
As a proof of concept, we have added support for the following tools: (1) PyCG, (2) \headergen, (3) Transformers, (4) Ollama, (5) TAJS, and (6) Jelly.

\swarmcg supports multiple programming languages, starting with Python and JavaScript, with ongoing efforts to integrate Java and plans to extend to additional languages.
The suite is designed to be community-driven, encouraging contributions from both static analysis experts and enthusiasts, making it a dynamic and evolving resource for the research community. 

\subsection{SWARM-JS: JavaScript Micro-benchmark}
Despite the increasing importance of JavaScript analysis, the availability of well-defined benchmarks tailored for JavaScript call-graph construction remains limited. 
Existing benchmarks, such as SunSpider~\citep{webkit_sunspider}, part of the WebKit browser engine, are primarily designed to test the performance aspects of JavaScript engines rather than facilitating program analysis.
SunSpider includes single-file JavaScript examples that represent real-world scenarios, but it does not provide explicit ground truth for call-graphs. 

In a recent study by \cite{Antal2023IsJavaScript}, the authors assessed static call-graph techniques using the SunSpider benchmark by manually comparing the call-graphs generated by the tools with the source code. 
Precision was measured by verifying whether specific edges in the graph were accurately identified. 
However, this manual approach limits the scope of the evaluation and limits the extensibility of the respective research.
Furthermore, the lack of attention to recall in this manual evaluation process results in an incomplete understanding of the tools' performance.

To address these limitations, we developed a new JavaScript micro-benchmark, SWARM-\textit{JS}, tailored specifically for call-graph construction.
Inspired by call-graph micro benchmarks in Python, such as PyCG and Jarvis, our benchmark aims to provide a systematic and comprehensive set of test cases that reflect the diverse language-specific constructs of JavaScript.

To construct the benchmark, we followed a methodology similar to that used by the authors of the PyCG~\citep{salisPyCGPracticalCall2021c} call-graph benchmark for Python.
Their process consists of three main steps:
(1) identifying a diverse set of language features relevant to call-graph construction,
(2) designing minimal test scripts inspired by real-world uses of these features, and
(3) conducting expert review to validate correctness and representativeness.

Applying this methodology to JavaScript, we began by surveying the ECMAScript specification~\citep{ecma-262} and the existing SunSpider~\citep{webkit_sunspider} JavaScript benchmark to identify essential language features and edge cases.
A comparative analysis with Python call-graph benchmarks, such as PyCG and Jarvis, helped determine which test scenarios could be adapted to JavaScript.
Test cases were constructed by re-implementing the intent of PyCG's benchmark scenarios in JavaScript while maintaining feature isolation.
For instance, Python’s lambda expressions were mapped to JavaScript’s arrow functions, given their similar semantics. 
In contrast, JavaScript-specific constructs such as prototypes, dynamic property access, and mixins were created additionally.

\textbf{Validity of SWARM-JS.}
To ensure correctness and reliability, all test cases and their ground truths were manually reviewed and refined through multiple iterations. 
A JavaScript expert independently validated a randomly selected subset of 25 test cases to verify the accuracy and correctness of the ground truth.
Based on the expert's feedback, we revised the benchmark to correct ground truth annotations. This iterative review process improved the overall validity of the benchmark.\\

The resulting benchmark, SWARM-JS, comprises 126 JavaScript code snippets, organized into 18 feature categories.
Table~\ref{tab:benchmark_categories} presents the complete list of categories along with the number of test cases and their descriptions. 
Each snippet in the benchmark is accompanied by a corresponding ground truth file, which provides the expected call-graph.
The ground truth schema follows the PyCG benchmark, allowing for a consistent framework for evaluating call-graph accuracy across different languages.
The code snippets and ground truth information were manually inspected and iteratively refined to ensure correctness.

An example code snippet is shown in Listing~\ref{lst:swarm_js_code} and its corresponding ground truth is given in Listing~\ref{lst:swarm_js_gt}.

\begin{table}[]
    \centering
	\renewcommand{\arraystretch}{1.2}
    
    \caption{Distribution of 126 JavaScript code snippets into 18 feature categories in SWARM-JS micro-benchmark}
    \label{tab:benchmark_categories}
    \begin{tabular}{lrp{7cm}}
        \toprule
        \textbf{Category} & \textbf{\# Cases} & \textbf{Description} \\
        \midrule
        Args & 10 & Positional and default argument passing. \\
        Assignments & 8 & Variable assignments and reassignments. \\
        Builtins & 3 & Built-in functions and objects. \\
        Classes & 21 & Class definitions, methods, and inheritance. \\
        Decorators & 7 & Use of function and class decorators. \\
        Objects & 12 & Object creation, property access, and manipulation. \\
        Direct Calls & 9 & Focuses on direct function and method calls. \\
        Dynamic & 1 & Dynamic code injection and method access. \\
        Exceptions & 3 & Exception handling. \\
        Functions & 4 & Function declarations and expressions. \\
        Generators & 6 & Generator functions and yield behavior. \\
        Imports & 15 & Module imports and exports. \\
        Kwargs & 3 & Keyword arguments and related constructs. \\
        Arrow Functions & 5 & Arrow function syntax and behavior. \\
        Arrays & 8 & Array manipulation and iteration. \\
        Inheritance & 4 & Prototype-based and class-based inheritance. \\
        Mixins & 3 & Mixin patterns for object composition. \\
        Returns & 4 & Functions that return other functions. \\
        \bottomrule
    \end{tabular}
\end{table}

\begin{minipage}[t]{.5\textwidth}
	\captionsetup[lstlisting]{singlelinecheck=off,justification=raggedright}
	\begin{lstlisting}[language=JavaScript, caption=Code snippet of main.js, style=pstyle, label=lst:swarm_js_code]
function paramFunc() {}


function func(a) {
    a();
}


func(paramFunc); 
\end{lstlisting}
\end{minipage}%
\begin{minipage}[t]{.5\textwidth} 
 \captionsetup[lstlisting]{singlelinecheck=off}
	\begin{lstlisting}[caption=Ground truth for main.js, language=json, label=lst:swarm_js_gt]
{
    "main.func": [
      "main.paramFunc"
    ],
    "main.paramFunc": [],
    "main": [
      "main.func"
    ]
}
\end{lstlisting}
\end{minipage}%

\subsection{TypeEvalPy Autogen Extension}
The micro-benchmark that is part of the \typeevalpy framework is constrained by its limited representation of Python base types, covering only 860 types derived from 154 code snippets. 
This limited coverage has implications in the context of evaluating LLMs.
Since a large proportion of type annotations in the \typeevalpy benchmark are str, LLMs might have exhibited high exact matches due to overrepresentation, rather than a genuine understanding of diverse type usages.
This narrow focus undermines the applicability and robustness of the evaluation results, as the models are not thoroughly tested on a wider variety of base types available in Python.

To address this limitation, we extend \typeevalpy by integrating auto-generation capabilities aimed at broadening the type diversity in the micro-benchmark. 
This enhancement is realized through a systematic process of template-based code generation. 
We first designed templates for the existing code snippets, introducing placeholders, such as $<value1>$, which are dynamically replaced by different types during code generation.
Additionally, associated configuration files were created to map these placeholders to various possible type values. 
For example, in Listing \ref{lst:typeevalpy_autogen}, the code snippet includes placeholders at specific locations, and the corresponding code generated is shown in Listing \ref{lst:typeevalpy_autogen_example}.
Type annotation ground-truth and values are generated based on the configuration rules outlined in Listing \ref{lst:typeevalpy_autogen_config}.
As an illustration, line 2 in Listing \ref{lst:typeevalpy_autogen} aligns with lines 15 and 31 in Listing \ref{lst:typeevalpy_autogen_config}, which define the relevant type mappings.

The auto-generation process systematically computes all permutations of types for the placeholders.
For example, with four configured types and two placeholders, the generator produces 12 unique programs based on the formula $P(n,r) = n! / (n-r)!$ where $n$ is the total number of configured types, and $r$ is the number of placeholders in a given template. 
Each program is generated with a unique arrangement of types, such as (str, float) and (str, int). This method enables the creation of a comprehensive range of programs with different type configurations, enhancing the diversity of the benchmark.
Note that the values are generated randomly for each of these placeholders according to their data types.
For instance, an example of the generated test case for this template is shown in Listing \ref{lst:typeevalpy_autogen_example} and its associated ground truth in Listing \ref{lst:typeevalpy_autogen_example_gt}.

Special cases, such as lists and dictionaries, require additional handling to ensure that every element within these data structures is correctly annotated. 
Similarly, imported code segments demand careful modelling to avoid inconsistencies in the generated programs. 
These complexities were addressed within the generator, which was carefully designed to ensure that all edge cases were correctly handled.

Once the programs are generated, each is executed to verify its correctness. 
If the program executes without errors, it is retained in the benchmark. 
In contrast, programs that fail due to type incompatibility, such as attempting to add a string to a float, are discarded.
This filtering ensures that only valid test cases are included in the final benchmark.

The creation of the auto-generated benchmark was a collaborative effort.
The first author was responsible for creating the initial templates, while the second author verified the generated programs for correctness, iteratively fixing errors and ensuring the accuracy of type annotations.
Furthermore, note that the programs have a single execution path, therefore avoiding ambiguities in the ground-truth.

The auto-generation capability expands the \typeevalpy benchmark with 7,121 Python files, containing a total of 77,268 type annotations.
This increase in both the quantity and variety of annotated types ensures a more comprehensive framework for evaluating the performance and generalizability of LLMs in type inference tasks.

\begin{minipage}[t]{.5\textwidth}
	\captionsetup[lstlisting]{singlelinecheck=off,justification=raggedright}
	\begin{lstlisting}[language=Python, caption=Template for main.py, style=pstyle, label=lst:typeevalpy_autogen]
def func1():
    return <value1>


def func2():
    return <value2>

a = func1()
b = func2()
\end{lstlisting}

\end{minipage}%
\begin{minipage}[t]{.5\textwidth}
	\captionsetup[lstlisting]{singlelinecheck=off,justification=raggedright}
	\begin{lstlisting}[language=Python, caption=Generated main.py, style=pstyle, label=lst:typeevalpy_autogen_example]
def func1():
    return 34


def func2():
    return 53.24

a = func1()
b = func2()
\end{lstlisting}

\end{minipage}

\begin{minipage}[b]{.55\textwidth}
 \captionsetup[lstlisting]{singlelinecheck=off}
	\begin{lstlisting}[caption=Autogen configuration\\ for main.py, language=json, label=lst:typeevalpy_autogen_config]
{
"replacement_mode": "Complex",
"type_replacements": [
    "int",
    "float",
    "str",
    "bool"
],
"ground_truth": [{
    "file": "main.py",
    "line_number": 1,
    "col_offset": 5,
    "function": "func1",
    "type": [
        "<value1>"
    ]},
    {
    "file": "main.py",
    "line_number": 5,
    "col_offset": 5,
    "function": "func2",
    "type": [
        "<value2>"
    ]},
    {
    "file": "main.py",
    "line_number": 8,
    "col_offset": 1,
    "variable": "a",
    "type": [
        "<value1>"
    ]},
    {
    "file": "main.py",
    "line_number": 9,
    "col_offset": 1,
    "variable": "a",
    "type": [
        "<value2>"
    ]}]
}
\end{lstlisting}
\end{minipage}%
\begin{minipage}[b]{.45\textwidth}
	\captionsetup[lstlisting]{singlelinecheck=off}
	\begin{lstlisting}[caption=Generated ground truth for main.py, language=json, label=lst:typeevalpy_autogen_example_gt]
[{
    "file": "main.py",
    "line_number": 1,
    "col_offset": 5,
    "function": "func1",
    "type": [
        "int"
    ]
},
{
    "file": "main.py",
    "line_number": 5,
    "col_offset": 5,
    "function": "func2",
    "type": [
        "float"
    ]
},
{
    "file": "main.py",
    "line_number": 8,
    "col_offset": 1,
    "variable": "a",
    "type": [
        "int"
    ]
},
{
    "file": "main.py",
    "line_number": 9,
    "col_offset": 1,
    "variable": "a",
    "type": [
        "float"
    ]
}]
\end{lstlisting}
\end{minipage}

\section{Methodology}
\label{sec:methodology}
We next describe the experimental setup, the model selection criteria, the prompt design, and the metrics used to investigate these RQs.

\subsection{Model Selection}
In this extension study, we selected LLMs for evaluation by focusing on organizations that are actively conducting research and releasing state-of-the-art models on the Hugging Face platform.\footnote{\url{https://huggingface.co/}}
We shortlisted five prominent organizations from Hugging Face that are building foundational models: Alibaba, Google, Meta, Microsoft, and Mistral.
Apart from the models with open weights, we chose OpenAI as the proprietary service provider to compare against open models. 

We selected a total of \textit{24} LLMs across all organizations.
From the organizations we shortlisted, we included all the instruction-tuned models, which are fine-tuned for following user instructions, across all available parameter sizes.
This included multiple variations of the models, such as 7B, 13B, and larger configurations, allowing for a comprehensive evaluation across different scales. 
In addition to general-purpose models, we also included specialized code models, which are optimized for code understanding and related tasks, as these models are expected to perform better on code-specific benchmarks. 

Two closed-source models from OpenAI, gpt-4o and gpt-4o-mini, were included due to their superior performance in general-purpose tasks, providing a benchmark for comparison against open-source models.
We limited the number of proprietary models we test to optimize costs and chose OpenAI's GPT models due to their popularity.

The list of models evaluated in this study is listed in the Table~\ref{tab:models}.

\begin{table}[]
    \centering
    \caption{Selected Models and Parameter Sizes}
    \label{tab:models}
    \begin{tabular}{llr}
        \toprule
        \multirow{2}{*}{\textbf{Organization}} & \multirow{2}{*}{\textbf{Model Name}} & \textbf{Parameter Size} \\
        & & \textbf{(billion)} \\
        \midrule
        \multirow{2}{*}{Alibaba} 
            & Qwen/Qwen2-7B-Instruct & 7B \\
            & Qwen/Qwen2-72B-Instruct & 72B \\
        \midrule
        \multirow{3}{*}{Google} 
            & google/gemma-2-2b-it & 2B \\
            & google/gemma-2-9b-it & 9B \\
            & google/gemma-2-27b-it & 27B \\
        \midrule
        \multirow{2}{*}{GPT} 
            & gpt-4o & - \\
            & gpt-4o-mini & - \\
        \midrule
        \multirow{6}{*}{Meta} 
            & meta-llama/CodeLlama-7b-Instruct-hf & 7B \\
            & meta-llama/CodeLlama-13b-Instruct-hf & 13B \\
            & meta-llama/CodeLlama-34b-Instruct-hf & 34B \\
            & meta-llama/Meta-Llama-3.1-8B-Instruct & 8B \\
            & meta-llama/Meta-Llama-3.1-70B-Instruct & 70B \\
            & TinyLlama/TinyLlama-1.1B-Chat-v1.0 & 1.1B \\
        \midrule
        \multirow{5}{*}{Microsoft} 
            & microsoft/Phi-3-small-128k-instruct & 7.3B \\
            & microsoft/Phi-3-medium-128k-instruct & 14B \\
            & microsoft/Phi-3.5-mini-instruct & 3.8B \\
            & microsoft/Phi-3.5-MoE-instruct & 41.9B \\
            & microsoft/Phi-3-mini-128k-instruct & 3.8B \\
        \midrule
        \multirow{6}{*}{Mistral} 
            & mistralai/Mixtral-8x22B-Instruct-v0.1 & 141B \\
            & mistralai/Mixtral-8x7B-Instruct-v0.1 & 46.7B \\
            & mistralai/Mistral-7B-Instruct-v0.3 & 7B \\
            & mistralai/Mistral-Nemo-Instruct-2407 & 12.2B \\
            & mistralai/Mistral-Large-Instruct-2407 & 123B \\
            & mistralai/Codestral-22B-v0.1 & 22B \\
        \bottomrule
    \end{tabular}
\end{table}

\subsection{Prompt Design}
To optimize prompt design, we adopted an iterative and experimental approach \citep{chenUnleashingPotentialPrompt2023, schulhoff_prompt_2024}.
Initial efforts focused on enhancing the prompt by including detailed task descriptions and specifying the expected response format.
Notably, we used a one-shot prompting technique, embedding an example question and answer within the prompt.
The one-shot prompt example was designed using the simplest program that encapsulates key aspects of the expected output for the given task. 
For instance, in the type inference task, the example included variables of different types to ensure variety.
The decision to use a simple example was primarily to ensure that the model's responses adhered to the desired format, enabling reliable parsing of the results.
Additionally, using a complex example in a one-shot setting does not always improve performance. 
Prior research by \cite{chenUnleashingPotentialPrompt2023} indicates that for sufficiently complex models, like those used in this study, a well-structured zero-shot prompt can be as effective as, or even preferable to, a complex few-shot prompt.

Despite these refinements, we encountered challenges with the LLM's ability to produce \emph{structured} outputs.
Our experiments revealed that even with explicit instructions to generate outputs in JSON format, models struggled to deliver results that could be reliably parsed.
To address this, we explored a question-answer based method, querying the model and then translating its natural-language responses back into a structured JSON format.

To further improve reliability, we analyzed the initial output to refine the prompt, particularly for cases where models failed to generate accurate results in response to a simple prompt.
For instance, we noted that the aliases of program variables were not being considered in the final output. Therefore, we introduced generic instructions to ensure alias tracking in the program. 
Note that the same prompt is used to evaluate all models, including code models.

In the following sections, we discuss the prompts for type inference and call-graphs tasks in detail. 

\subsubsection{Type-Inference Prompts}

The prompt design employed in this study follows a structured two-part approach to guide the LLM through the task. 
The first part provides a detailed description of the task necessary to conduct the analysis, ensuring clarity in the expected operations. 
This is followed by the second part, which includes an example input-output pair in line with the one-shot prompting technique.
Additionally, instructions on the format of the output are explicitly provided to direct the model’s responses.
Finally, the code relevant to the task is added to the prompt.
Note that for test cases with file imports, all the relevant file contents are added to the prompt with relative file names to indicate the file structure of the test case.

Despite the careful structuring, we encountered difficulties in the initial attempts to generate valid JSON output using this approach.
Specifically, the model often failed to consistently produce JSON in the required format. 
The primary issues observed were missing keys or the inclusion of unexpected keys, attributed to the LLM’s inability to adhere to the complex output schema.
The underlying complexity of the type annotations schema of the \typeevalpy framework presented additional challenges for LLMs.

To address these limitations, the task complexity was reduced by breaking the task down into a series of \textit{question-answer} pairs and using the one-shot prompting technique.
This approach simplifies the requirement to follow specific output schema and enhances its ability to follow the prompt more accurately.
For example, as shown in Listing \ref{lst:prompt_qa_types_2}, three specific questions were generated based on the variables declared in the one-shot code example.
These questions include the name of the variable and the location of its declaration.
Additionally, placeholders were introduced for each question, with sequential numbers to indicate where the model should provide responses.

The actual questions were generated using ground-truth data. 
By iterating over the variables and functions listed in the ground-truth, appropriate questions were formulated. 
However, in a practical setting, this task could be automated using the program’s abstract syntax tree (AST). For this study, the available ground truth data was used to simplify the implementation.

To clarify, consider the full prompt in Listing \ref{lst:actual_prompt}.
In this case, from the ground-truth, we know that five program identifiers require type inference.
The five questions in the prompt are generated by iterating over the ground-truth data and extracting the identifier names, along with their corresponding line and column numbers. 
In theory, this information could be obtained by parsing the AST of the program.

Finally, the model’s responses were parsed using regular expressions, which enabled the correct mapping of answers back to the original questions. 
This method allowed for generating JSON outputs that adhered to the \typeevalpy schema, which were then used for the evaluation.
To demonstrate this in practice, we have listed an example with the source code, ground-truth, model response, and parsed JSON in Listing \ref{sec:response_mistral_types}.

\subsubsection{Call-Graph Prompts}

The design of prompts for call-graph analysis follows an approach similar to the one described in the previous section. 
Initially, a detailed description of the task is provided, which is followed by an example input-output pair according to the target language.
The task description outlines the specific requirements for analyzing the call-graph, and instructions for formatting the output are included to ensure consistency in the model’s responses, as shown in Listing~\ref{lst:prompt_qa_cg_1}.

To generate questions within these prompts, a method akin to the one used previously is used.
The first question typically addresses function calls at the module level, followed by questions regarding each individual call made within function definitions as illustrated in Listing~\ref{lst:prompt_qa_cg_2}.

In practical scenarios, these questions can be generated by iterating through the AST of the program. By identifying function definitions and call nodes within the AST, the necessary information can be extracted. 
However, for this study, ground truth data was used to formulate the questions, allowing for a more straightforward implementation.

Additionally, for flow-sensitive call-graph analysis, the prompts were adjusted to accommodate the location of the call site.
Listings \ref{lst:prompt_qa_cs_1} and \ref{lst:prompt_qa_cs_2} present the specific prompts used for constructing flow-sensitive call-graphs.

\textbf{Note on Context Length.}
The maximum prompt size encountered across both the call-graph and type inference benchmarks was 1,287 tokens, as measured by the gpt-4o tokenizer.
This means that the prompts used in this study were comfortably within the context limits of all the LLMs evaluated.
The model with the largest context size, gpt-4o, supports up to 128,000 tokens, while the smallest context size was offered by TinyLlama-1.1b, which has a limit of 2,048 tokens.

For reference, the cumulative size of the prompts from the entire \typeevalpy micro-benchmark amounts to 69,563 tokens.
Even in this case, the total prompt size remains well below the maximum context length of most models evaluated, ensuring that the models had enough capacity to process the full input without truncation.

\subsection{Evaluation Metrics}
In this study, we measured completeness, soundness, and exact matches to assess both flow-insensitive callgraph construction and flow-sensitive call-site extraction.
Furthermore, we use the exact matches metric to evaluate type inference performance.

\textbf{Completeness and Soundness.} 
In this study, we use the terms completeness and soundness as they have been pre-established in call-graph research \citep{salisPyCGPracticalCall2021c, venkatesh_static_2024}. The terms completeness and soundness are closely related to the precision and recall metrics. 

Precision is directly tied to completeness, as it measures the proportion of correctly identified call edges relative to all edges produced by the model. A complete call-graph will have perfect precision, as it contains no false positives. This terminology can be a bit confusing at first because it implies that a call-graph that is ``incomplete'' in the above sense is not one that misses call edges but one that has spurious edges. The reader shall keep that in mind.

Recall is closely related to soundness, as it measures the proportion of true call edges that are correctly identified. A sound call-graph will demonstrate perfect recall by including all true call edges, without omitting any.

Here, completeness and soundness are measured at the individual test case level within the benchmark.
A test case is considered complete if there are no false positives in the generated call-graph for that specific case. 
Similarly, it is considered sound if there are no false negatives. 
This means that if even a single false positive or false negative is detected in the responses generated for a test case, it is marked as a failure in terms of completeness or soundness, respectively.

However, precision and recall have specific implications when evaluated at the level of individual test cases, particularly in a micro-benchmark setting.
Rather than measuring how precise or recall-efficient a system is overall, it is more insightful to determine whether a test case is fully complete or sound with respect to the specific feature being tested.
This binary evaluation, either complete or sound, provides clearer insights into whether specific features are fully captured, without the ambiguity that partial correctness metrics like precision or recall might introduce.
This evaluation approach mirrors the methodologies used in previous studies, specifically in \pycg~\citep{salisPyCGPracticalCall2021c} and \headergen~\citep{venkateshEnhancingComprehensionNavigation2023a}. 

\textbf{Exact matches.} 
The exact-matches metric for the call-graph measures the number of function calls that exactly match the ground truth.
To compute this, we compare the expected calls for each node in the ground truth with those produced by the model. 
For nodes where both lists are non-empty, we count exact matches when every element in the generated list appears in the ground truth.
For nodes with empty lists, an exact match is counted if the model also produces an empty list.

Furthermore, aligning with the literature \citep{Typilus, mirType4PyPracticalDeep2022c, HiTyper, venkateshTypeEvalPyMicrobenchmarkingFramework2023, venkatesh_emergence_2024}, for type-inference evaluation, we use exact matches as the metric as well.

\textbf{Time.}
Time measurements were taken on open models, as they were all executed on the same hardware using identical parameters for model loading and inference.
To ensure uniformity in the testing setup, all models were loaded using 4-bit quantization, with a batching size of 12.
To ensure a fair comparison, we applied the same batching size across all models.
While smaller models could, in practical scenarios, process more prompts per batch due to lower memory requirements, we chose to standardize the testing conditions.
This approach prevents smaller models from having an advantage and allows for a fair assessment.

The time recorded represents the total time needed to process all benchmark test cases. 
Time measurements for OpenAI models were omitted, as they were inferred using a batch API that returns results after 24 hours at a 50\% lower cost.
Given that these models were not run on our hardware, a direct comparison with the open models would not be appropriate.

\subsection{Implementation Details}

For the implementation of our experiments, we used the Hugging Face transformers~\citep{wolf_transformers_2020} Python interface to run LLMs on our hardware.
This interface provides a flexible and efficient environment to manage inference tasks across multiple models.
The models were loaded using 4-bit quantization, with a batch size of 12, and configured to use greedy search.
Greedy search was chosen to always select the most probable next token, ensuring deterministic outputs across all runs.

To conduct the type-inference experiments, we extended the existing \typeevalpy framework. This allowed for seamless integration with our testing pipeline.
For the call-graph experiments, we built a custom adaptor within the \swarmcg framework.

All experiments were run on the following hardware configuration: one NVIDIA H100-80GB GPU, 16 Intel(R) Xeon(R) Platinum 8462Y+ processors, and 78 GB of memory.

\textbf{Note on Quantization.}
To optimize resource utilization, we chose to load models using 4-bit quantization, allowing large models to be efficiently deployed on a single H100 GPU with 80GB of memory. 
This approach significantly reduces computational and memory requirements while maintaining the feasibility of running extensive experiments.
Furthermore, prior research indicates that quantization has minimal impact on the accuracy of large models, making it a viable strategy for balancing efficiency and performance \citep{langComprehensiveStudyQuantization2024, jinComprehensiveEvaluationQuantization2024, dettmersQLoRAEfficientFinetuning2023a}.
\section{Results}
\label{sec:results}

\noindent
We next address the research questions and highlight the key results from our different analysis.

\subsection{RQ1: Accuracy of Callgraph Analysis}

The results of our experiments for flow-insensitive call-graph analysis for Python and JavaScript are presented in Tables \ref{tab:cg_py} and \ref{tab:cg_js}, respectively. 
The Python results are based on the PyCG micro-benchmark suite, while the JavaScript results use the SWARM-JS micro-benchmark.
Additionally, Table \ref{tab:cs} provides the results for flow-sensitive call-graph analysis based on the \headergen micro-benchmark. 
In the following sections, we discuss each of these results in detail.

\subsubsection{Flow-insensitive Call-graph Analysis}

\begin{table}[t]
    \centering
	\renewcommand{\arraystretch}{1.3}
	\setlength{\tabcolsep}{5.5pt} % Adjusts the space between columns
	\caption{Comparative analysis across LLMs for \textbf{flow-insensitive} call-graph analysis on the \pycg Python micro-benchmark}
	\label{tab:cg_py}
\begin{tabular}{lrrrr}
\multicolumn{5}{r}{\priority{100} 80-100\%,
\priority{70} 60-80\%,
\priority{50} 40-60\%,
\priority{30} 20-40\%,
\priority{10} 0-20\%}  \\  
\toprule
\multirow{2}{*}{\centering \textbf{Model}} & \multicolumn{1}{c}{\textbf{Complete}} & \multicolumn{1}{c}{\textbf{Sound}} & \multicolumn{1}{c}{\textbf{Exact Matches}} & \multirow{2}{*}{\textbf{Time (s)}} \\ 
\cmidrule(lr){2-3} \cmidrule(lr){4-4}
 & \multicolumn{2}{c}{126 cases} & \multicolumn{1}{c}{599 cases} & \\
\midrule
\rowcolor{gray!20} \textbf{PyCG} & \textbf{107} \priority{85} & \textbf{110} \priority{87} & \textbf{569} \priority{95} & \textbf{0.41} \\
\rowcolor{gray!20} \textbf{mistral-large-it-2407-123b} & 76 \priority{61} & 79 \priority{62} & 518 \priority{86} & 534.01 \\
gpt-4o & 63 \priority{50} & 75 \priority{59} & 486 \priority{81} & n/a \\
qwen2-it-72b & 35 \priority{28} & 67 \priority{53} & 427 \priority{71} & 286.17 \\
llama3.1-it-70b & 37 \priority{29} & 63 \priority{51} & 424 \priority{71} & 277.69 \\
gpt-4o-mini & 47 \priority{37} & 47 \priority{37} & 397 \priority{66} & n/a \\
mistral-nemo-it-2407-12.2b & 50 \priority{39} & 44 \priority{35} & 358 \priority{59} & \textbf{59.35} \\
gemma2-it-27b & 29 \priority{23} & 43 \priority{34} & 344 \priority{57} & 180.94 \\
llama3.1-it-8b & 2 \priority{2} & 40 \priority{32} & 179 \priority{30} & 185.52 \\
\rowcolor{gray!20} \textbf{mixtral-v0.1-it-8x22b} & \textbf{2} \priority{2} & 65 \priority{51} & \textbf{173} \priority{29} & 1106.46 \\
\rowcolor{gray!20} \textbf{phi3.5-moe-it-41.9b} & 9 \priority{7} & 25 \priority{19} & 166 \priority{28} & \textbf{3335.89} \\
phi3-small-it-7.3b & 3 \priority{2} & 10 \priority{8} & 150 \priority{25} & 73.92 \\
phi3-medium-it-14b & 3 \priority{2} & 29 \priority{23} & 142 \priority{24} & 140.55 \\
codellama-it-34b & 64 \priority{51} & 23 \priority{18} & 130 \priority{22} & 545.96 \\
qwen2-it-7b & 2 \priority{2} & 42 \priority{33} & 128 \priority{21} & 52.74 \\
mistral-v0.3-it-7b & 4 \priority{3} & 25 \priority{21} & 122 \priority{21} & 74.74 \\
codestral-v0.1-22b & 34 \priority{27} & 24 \priority{19} & 120 \priority{21} & 582.78 \\
tinyllama-1.1b & 35 \priority{28} & 5 \priority{4} & 91 \priority{15} & 507.21 \\
\rowcolor{gray!20}  \textbf{gemma2-it-9b} & \textbf{120} \priority{95} & 4 \priority{3} & \textbf{63} \priority{10} & \textbf{1587.85} \\
mixtral-v0.1-it-8x7b & 9 \priority{7} & 19 \priority{15} & 52 \priority{9} & 1221.64 \\
codellama-it-13b & 20 \priority{16} & 17 \priority{13} & 43 \priority{7} & 619.04 \\
codellama-it-7b & 7 \priority{6} & 4 \priority{3} & 31 \priority{5} & 281.14 \\
phi3-mini-it-3.8b & 1 \priority{1} & 7 \priority{6} & 24 \priority{4} & 85.48 \\
\rowcolor{gray!20} \textbf{gemma2-it-2b} & \textbf{117} \priority{93} & 1 \priority{1} & \textbf{10} \priority{2} & \textbf{731.7} \\
phi3.5-mini-it-3.8b & 2 \priority{2} & 6 \priority{5} & 9 \priority{2} & 81.3 \\

\bottomrule                       
\end{tabular}
\end{table}

This section reports the results obtained in the \textbf{flow-insensitive call-graph analysis} for Python and JavaScript programs, separately.

\textbf{Python Programs.} The results of our evaluation, presented in Table \ref{tab:cg_py}, highlight the superior performance of the static analysis tool PyCG compared to LLMs in terms of completeness, soundness, exact matches, and processing time.
Specific rows and values that are discussed in the text are highlighted in the table for clarity.

In a benchmark of 126 test cases, PyCG achieved 84.9\% completeness and 87.3\% soundness. 
This means that for the majority of test cases, PyCG produced no false positives (completeness) and missed very few valid function calls (soundness). 
These results significantly surpass those of the closest competing model, mistral-large-it-2407-123b, which attained 60.3\% completeness and 62.6\% soundness. 
Additionally, PyCG produced 569 exact matches out of 599, outperforming mistral-large-it-2407-123b by 51 matches.

The model mistral-large-it-2407-123b shows moderate performance in both completeness and soundness. 
However, it achieved a high exact match score of 86.4\%, indicating that while it correctly identifies many function-call relations, it also introduces false positives and misses valid ones.
This leads to failures in both completeness and soundness across many test cases, suggesting that the model lacks support for certain Python language features.

gpt-4o ranks third, performing behind mistral-large-it-2407-123b, which suggests that open-source models may be catching up to closed-source models. However, most other open-source models underperformed significantly.

The model gemma2-it-9b displayed a notable discrepancy between completeness (95\%) and soundness (3\%), suggesting that while it rarely introduces false positives, it misses a vast number of valid function calls, leading to numerous test cases failing the soundness criterion.
The poor exact match score of 10.5\% reflects this imbalance.
Furthermore, its runtime of 1587 seconds makes it surprising given that the model is relatively small with 9 billion parameters.

The poor performance of mixtral-v0.1-it-8x22b, especially for a model with 141 billion parameters, demonstrates its limitations in handling the test cases. On the contrary, tinyllama-1.1b, despite being a smaller model, took significant time to process and performed poorly across all metrics.

To clarify how the results are parsed and evaluated, we provide an example in the appendix, showcasing the source code, ground truth, raw LLM response, and parsed call-graph JSON for both the top-performing model, mistral-large-it-2407-123b, and the least-performing model, phi3.5-mini-it-3.8b, for the same test case in our benchmark. These examples can be found in Sections \ref{sec:response_mistral} and \ref{sec:response_phi}, respectively.

% \subsubsection{JavaScript flow-insensitive Call-graph Analysis}

\begin{table}[]
    \centering
	\renewcommand{\arraystretch}{1.3}
	\caption{Comparative analysis across LLMs for \textbf{flow-insensitive} call-graph analysis on the SWARM-JS JavaScript micro-benchmark}
	\label{tab:cg_js}

\begin{tabular}{@{}lrrrr@{}}
\multicolumn{5}{r}{\priority{100} 80-100\%,
\priority{70} 60-80\%,
\priority{50} 40-60\%,
\priority{30} 20-40\%,
\priority{10} 0-20\%}  \\  
\toprule
\multirow{2}{*}{\centering \textbf{Model}} & \multicolumn{1}{c}{\textbf{Complete}} & \multicolumn{1}{c}{\textbf{Sound}} & \multicolumn{1}{c}{\textbf{Exact Matches}} & \multirow{2}{*}{\textbf{Time (s)}} \\ 
\cmidrule(lr){2-3} \cmidrule(lr){4-4}
 & \multicolumn{2}{c}{126 cases} & \multicolumn{1}{c}{596 cases} & \\ 
\midrule
\rowcolor{gray!20} \textbf{Jelly} & \textbf{49} \priority{39}& \textbf{85} \priority{67}& \textbf{490} \priority{82}& 643.51\\
\rowcolor{gray!20} \textbf{mistral-large-it-2407-123b}& \textbf{51} \priority{41}& \textbf{54} \priority{43}& \textbf{458} \priority{77}& 537.86\\
\rowcolor{gray!20} \textbf{gpt-4o} & 44 \priority{35}& \textbf{64} \priority{51}& 451 \priority{75}& n/a\\
qwen2-it-72b& 24 \priority{19}& 51 \priority{41}& 406 \priority{68}& 367.87\\
llama3.1-it-70b& 28 \priority{22}& 34 \priority{27}& 357 \priority{61}& 251.91\\
gpt-4o-mini& 32 \priority{25}& 30 \priority{24}& 347 \priority{58}& n/a\\
mistral-nemo-it-2407-12.2b& 44 \priority{35}& 23 \priority{18}& 311 \priority{52}& 56.5\\
gemma2-it-27b& 14 \priority{11}& 14 \priority{11}& 280 \priority{47}& 188.88\\
llama3.1-it-8b& 4 \priority{3}& 36 \priority{29}& 212 \priority{35}& 85.28\\
codestral-v0.1-22b& 35 \priority{28}& 19 \priority{15}& 145 \priority{24}& 496.08\\
\rowcolor{gray!20} \textbf{phi3.5-moe-it-41.9b} & 3 \priority{2}& 12 \priority{10}& 145 \priority{24}& \textbf{3344.09}\\
phi3-small-it-7.3b& 2 \priority{2}& 2 \priority{2}& 141 \priority{23}& 83.26\\
mistral-v0.3-it-7b& 6 \priority{5}& 19 \priority{15}& 132 \priority{22}& 1297.24\\
phi3-medium-it-14b& 2 \priority{2}& 17 \priority{13}& 130 \priority{22}& 145.18\\
codellama-it-34b& 74 \priority{59}& 14 \priority{11}& 103 \priority{17}& 540.06\\
mixtral-v0.1-it-8x22b& 2 \priority{2}& 33 \priority{26}& 94 \priority{16}& 90.21\\
qwen2-it-7b& 2 \priority{2}& 26 \priority{21}& 87 \priority{14}& 43.3\\
\rowcolor{gray!20} \textbf{TAJS}& \textbf{119} \priority{94}& 14 \priority{11}& 83 \priority{13}& \textbf{135.65}\\
\rowcolor{gray!20} \textbf{gemma2-it-9b}& \textbf{118} \priority{94}& 2 \priority{2}& 54 \priority{9}& \textbf{1593.44}\\
phi3-mini-it-3.8b& 2 \priority{2}& 2 \priority{2}& 40 \priority{7}& 77.61\\
tinyllama-1.1b& 14 \priority{11}& 1 \priority{1}& 37 \priority{6}& 499.58\\
codellama-it-7b& 14 \priority{11}& 0 \priority{0}& 30 \priority{5}& 272.88\\
codellama-it-13b& 3 \priority{2}& 9 \priority{7}& 21 \priority{4}& 598.55\\
phi3.5-mini-it-3.8b& 4 \priority{3}& 3 \priority{2}& 20 \priority{3}& 89.2\\
\rowcolor{gray!20} \textbf{gemma2-it-2b} & \textbf{116} \priority{92}& 1 \priority{1}& 13 \priority{2}& \textbf{721.01}\\
mixtral-v0.1-it-8x7b& 3 \priority{2}& 1 \priority{1}& 8 \priority{1}& 578.66\\
\bottomrule                           
\end{tabular}

\end{table}

\textbf{JavaScript Programs.} The results from analyzing the JavaScript benchmark (SWARM-JS) are presented in Table~\ref{tab:cg_js}.

Jelly, a hybrid static analysis tool, demonstrated strong performance in our evaluation. 
When executed with its approximate interpretation feature enabled~\citep{laursen_reducing_2024}, Jelly achieved a completeness score of 38.8\%, soundness of 67.4\%, and a high exact match rate of 82.2\%.
This configuration incorporates dynamic execution hints into the static analysis process, improving the tool’s ability to resolve function calls accurately.

In contrast, TAJS, which was previously shown to perform well in call-graph generation by~\citet{Antal2023IsJavaScript}, performed poorly in our evaluation. 
Although its results appeared to exhibit a low false-positive rate, further inspection revealed that this was due to widespread failures: 102 out of 126 SWARM-JS test cases resulted in analysis errors and produced empty outputs.
This is because TAJS only supports ECMAScript 3rd edition, whereas the SWARM-JS benchmark includes features from ECMAScript 6th edition, such as classes and arrow functions. 

The mistral-large-it-2407-123b model achieved 40.4\% completeness and 42.8\% soundness, making it the top-performing model overall, with an exact match score of 76.8\%.
Its runtime of 537.86 seconds, while not the fastest, is expected for a model of its size.
gpt-4o achieved 34.9\% completeness and 50.7\% soundness, with a total of
451 exact matches, performing closely to mistral-large-it-2407-123b for capturing valid function calls.

Models gemma2-it-9b and gemma2-it-2b showed high completeness scores (118 and 116, respectively), but very low soundness (2 and 1, respectively).
This indicates that although these models generated few false positives, they missed nearly all valid function calls, leading to largely empty call-graphs.
Furthermore, gemma2-it-9b had a very high runtime of 1593.44 seconds, making it both inefficient and ineffective.

\begin{table}[b]
    \centering
	\renewcommand{\arraystretch}{1.3}
	\setlength{\tabcolsep}{3.5pt} % Adjusts the space between columns
	\caption{Percentage comparison of models across Python and JavaScript \textbf{flow-insensitive call-graph} evaluations}
	\label{tab:cg_percentage_comparison}
\begin{tabular}{lrrr|rrr}
\multicolumn{7}{r}{\priority{100} 80-100\%,
\priority{70} 60-80\%,
\priority{50} 40-60\%,
\priority{30} 20-40\%,
\priority{10} 0-20\%}  \\  
\toprule
\multirow{2}{*}{\textbf{Model}} & \multicolumn{3}{c|}{\textbf{Python (\%)}} & \multicolumn{3}{c}{\textbf{JavaScript (\%)}} \\
\cmidrule(lr){2-4} \cmidrule(lr){5-7}
& \textbf{Comp.} & \textbf{Sound} & \textbf{EM.} & \textbf{Comp.} & \textbf{Sound} & \textbf{EM.} \\
\midrule
mistral-large-it-2407-123b & 60.32 \priority{61} & 62.70 \priority{63} & 86.64 \priority{87} & 40.48 \priority{41} & 42.86 \priority{43} & 76.85 \priority{77} \\
gpt-4o                    & 50.00 \priority{50} & 59.52 \priority{59} & 81.14 \priority{81} & 34.92 \priority{35} & 50.79 \priority{51} & 75.67 \priority{75} \\
qwen2-it-72b               & 27.78 \priority{28} & 53.17 \priority{53} & 71.29 \priority{71} & 19.05 \priority{19} & 40.48 \priority{41} & 68.12 \priority{68} \\
llama3.1-it-70b            & 29.37 \priority{29} & 50.00 \priority{50} & 70.78 \priority{71} & 22.22 \priority{22} & 26.98 \priority{27} & 59.90 \priority{59} \\
gpt-4o-mini                & 37.30 \priority{37} & 37.30 \priority{37} & 66.28 \priority{66} & 25.40 \priority{25} & 23.81 \priority{24} & 58.22 \priority{58} \\
mistral-nemo-it-2407-12.2b & 39.68 \priority{39} & 34.92 \priority{35} & 59.93 \priority{59} & 34.92 \priority{35} & 18.25 \priority{18} & 52.18 \priority{52} \\
gemma2-it-27b              & 23.02 \priority{23} & 34.13 \priority{34} & 57.43 \priority{57} & 11.11 \priority{11} & 11.11 \priority{11} & 46.98 \priority{47} \\
llama3.1-it-8b             &  1.59 \priority{2}  & 31.75 \priority{32} & 29.88 \priority{30} &  3.17 \priority{3}  & 28.57 \priority{29} & 35.57 \priority{36} \\
mixtral-v0.1-it-8x22b        &  1.58 \priority{2}  & 51.58 \priority{52} & 28.88 \priority{29} &  1.58 \priority{2}  &  26.19 \priority{26} & 15.77 \priority{16} \\
phi3.5-moe-it-41.9b        &  7.14 \priority{7}  & 19.84 \priority{19} & 27.71 \priority{28} &  2.38 \priority{2}  &  9.52 \priority{10} & 24.33 \priority{24} \\
\bottomrule
\end{tabular}
\end{table}

\textbf{Comparative Analysis of Python and JavaScript Results.}
In this section, we discuss the performance of LLMs across flow-insensitive call-graph evaluation for Python and JavaScript.
Table \ref{tab:cg_percentage_comparison}
compares the top 10 performing LLMs based on exact match rates for Python and JavaScript programs. Overall, the models exhibited stronger performance in Python than in JavaScript programs. The leading model, mistral-large-it-2407-123b, achieved an exact match rate of 86.6\% in Python, outperforming its JavaScript results, where it reached 76.8\%. This performance gap is consistent across other models, all of which show a noticeable decline in accuracy across metrics when evaluated on JavaScript.

\subsubsection{Flow-sensitive Call-graph Analysis}

This section reports the results obtained in the \textbf{flow-sensitive call-graph analysis} for Python programs.

\textbf{Python Programs.} Table \ref{tab:cs} presents the results of flow-sensitive call-graph analysis on the \headergen micro-benchmark, comparing the performance of various LLMs and the static analysis tool \headergen.

\begin{table}[t]
    \centering
	\renewcommand{\arraystretch}{1.3}
	\setlength{\tabcolsep}{4.5pt} % Adjusts the space between columns
	\caption{Comparative analysis across LLMs for \textbf{flow-sensitive} call-graph analysis on the \headergen micro-benchmark}
	\label{tab:cs}
\begin{tabular}{@{}lrrrr@{}}
\multicolumn{5}{r}{\priority{100} 80-100\%,
\priority{70} 60-80\%,
\priority{50} 40-60\%,
\priority{30} 20-40\%,
\priority{10} 0-20\%}  \\  
\toprule
\multirow{2}{*}{\centering \textbf{Model}} & \multicolumn{1}{c}{\textbf{Complete}} & \multicolumn{1}{c}{\textbf{Sound}} & \multicolumn{1}{c}{\textbf{Exact Matches}} & \multirow{2}{*}{\textbf{Time (s)}} \\ 
\cmidrule(lr){2-3} \cmidrule(lr){4-4}
 & \multicolumn{2}{c}{122 cases} & \multicolumn{1}{c}{357 cases} & \\ \midrule
\rowcolor{gray!20} \textbf{\headergen} & \textbf{111} \priority{91}& \textbf{112} \priority{92}& \textbf{326} \priority{91}& \textbf{10.94}\\
\rowcolor{gray!20} \textbf{mistral-large-it-2407-123b}& 38 \priority{31}& 32 \priority{26}& 102 \priority{29}& \textbf{581.87}\\
mixtral-v0.1-it-8x22b& 23 \priority{19}& 21 \priority{17}& 75 \priority{21}& 1123.6\\
\rowcolor{gray!20} \textbf{gpt-4o}& 31 \priority{25}& 26 \priority{21}& \textbf{74} \priority{21}& n/a\\
qwen2-it-72b& 22 \priority{18}& 19 \priority{16}& 70 \priority{21}& 289.96\\
codestral-v0.1-22b& 14 \priority{11}& 13 \priority{10}& 65 \priority{18}& 369.05\\
gemma2-it-27b& 22 \priority{18}& 15 \priority{12}& 55 \priority{15}& 173.85\\
llama3.1-it-70b& 19 \priority{16}& 17 \priority{14}& 54 \priority{15}& 362.55\\
\rowcolor{gray!20} \textbf{gpt-4o-mini}& 17 \priority{14}& 12 \priority{10}& \textbf{36} \priority{10}& n/a\\
mixtral-v0.1-it-8x7b& 11 \priority{9}& 11 \priority{9}& 32 \priority{9}& 1000.04\\
llama3.1-it-8b& 14 \priority{11}& 11 \priority{9}& 31 \priority{9}& 356\\
phi3-medium-it-14b& 10 \priority{8}& 9 \priority{7}& 31 \priority{9}& 131.57\\
phi3-mini-it-3.8b& 14 \priority{11}& 11 \priority{9}& 31 \priority{9}& 63.38\\
mistral-nemo-it-2407-12.2b& 18 \priority{15}& 11 \priority{9}& 26 \priority{7}& 54.06\\
phi3.5-mini-it-3.8b& 8 \priority{7}& 7 \priority{6}& 21 \priority{6}& 296.41\\
codellama-it-34b& 14 \priority{11}& 10 \priority{8}& 18 \priority{5}& 320.42\\
qwen2-it-7b& 11 \priority{9}& 9 \priority{7}& 16 \priority{4}& 48.82\\
mistral-v0.3-it-7b& 7 \priority{6}& 7 \priority{6}& 15 \priority{4}& 58.1\\
tinyllama-1.1b& 11 \priority{9}& 8 \priority{7}& 12 \priority{3}& 99.89\\
phi3-small-it-7.3b& 9 \priority{7}& 8 \priority{7}& 11 \priority{3}& 324.57\\
phi3.5-moe-it-41.9b& 8 \priority{7}& 6 \priority{5}& 8 \priority{2}& 2771.88\\
codellama-it-13b& 8 \priority{7}& 7 \priority{6}& 6 \priority{2}& 256.57\\
gemma2-it-9b& 6 \priority{5}& 6 \priority{5}& 5 \priority{1}& 1558.06\\
codellama-it-7b& 6 \priority{5}& 6 \priority{5}& 0 \priority{0}& 222.17\\
gemma2-it-2b& 6 \priority{5}& 6 \priority{5}& 0 \priority{0}& 716.7\\ \bottomrule
\end{tabular}
\end{table}

\headergen outperforms LLMs by achieving a completeness of 90.9\% and soundness of 91.8\%.
\headergen had 326 exact matches out of 357, achieving a score of 91.3\%.
This demonstrates that \headergen ensures low false positives and false negatives rates.
Additionally, it achieves this in 10.94 seconds, highlighting its efficiency compared to LLMs.

Among the LLMs, mistral-large-it-2407-123b stands out as the best-performing model, although it still falls significantly short of \headergen.
It achieved 31.1\% completeness and 26.2\% soundness, with 38 complete and 32 sound cases. 
Its 28.5\% exact match score (102 out of 357 cases) further highlights its limitations in capturing all function calls correctly.

All the other models underperformed in every metric, indicating that they failed to accurately capture the majority of function calls in the benchmark.  
When comparing these results to flow-insensitive analysis, the performance of LLMs further deteriorates. 
The increased complexity of flow-sensitive analysis, which requires specificity about the location of function calls, poses additional challenges for LLMs.
This seems to significantly reduce their ability to capture correct relationships, further highlighting the limitations of LLMs in handling more complex, context-specific analysis tasks.

\begin{tcolorbox}
Static analysis tools such as \pycg, Jelly, and \headergen generally outperformed LLMs in flow-insensitive and flow-sensitive call-graph analysis for both Python and JavaScript.
While \texttt{mistral-large} was the top-performing LLM, its overall accuracy consistently fell short of dedicated static analysis tools, highlighting the challenges LLMs face with complex code analysis.
\end{tcolorbox}

\subsection{RQ2: Accuracy of Type Inference}
\label{subsec:type-inf-res}

\subsubsection{TypeEvalPy Micro-benchmark}

\begin{table}[]
	\caption{Exact match comparison of LLMs for \textbf{type inference} on \typeevalpy micro-benchmark}
	\label{tab:type_inference_micro}
	\renewcommand{\arraystretch}{1.2}
	
	\begin{tabular}{lrrrrr}
   \multicolumn{6}{r}{\priority{100} 80-100\%,
\priority{70} 60-80\%,
\priority{50} 40-60\%,
\priority{30} 20-40\%,
\priority{10} 0-20\%}  \\  
		\multicolumn{6}{r}{ \textbf{FRT:} Function return type, \textbf{FPT:} Function parameter type, \textbf{LVT:} Local variable type}   \\
		\toprule
		\multicolumn{1}{c}{\textbf{Model}} & \multicolumn{1}{c}{\textbf{FRT}} & \multicolumn{1}{c}{\textbf{FPT}} & \multicolumn{1}{c}{\textbf{LVT}} & \multicolumn{1}{c}{\textbf{Total}} & \multirow{2}{*}{\textbf{Time (s)}}\\
\cmidrule(lr){2-2} \cmidrule(lr){3-3} \cmidrule(lr){4-4} \cmidrule(lr){5-5} 
		\multicolumn{1}{c}{Total} & \multicolumn{1}{c}{239} & \multicolumn{1}{c}{88} & \multicolumn{1}{c}{533} & \multicolumn{1}{c}{860} & \\       
		\midrule
\rowcolor{gray!20} \textbf{gpt-4o}                            & \textbf{217} \priority{91} & \textbf{81} \priority{92} & \textbf{508} \priority{95} & \textbf{806} \priority{94} & n/a                      \\
\rowcolor{gray!20} \textbf{mistral-large-it-2407-123b}        & 222 \priority{93} & 80 \priority{91} & 502 \priority{94} & \textbf{804} \priority{93} & 626.46                   \\
gpt-4o-mini                       & 212 \priority{89} & 80 \priority{91} & 491 \priority{92} & 783 \priority{91} & n/a                      \\
llama3.1-it-70b                   & 215 \priority{90} & 70 \priority{80} & 485 \priority{91} & 770 \priority{90} & 279.73                   \\
codestral-v0.1-22b                & 209 \priority{87} & 82 \priority{93} & 469 \priority{88} & 760 \priority{88} & 242.05                   \\
mixtral-v0.1-it-8x22b             & 196 \priority{82} & 68 \priority{77} & 492 \priority{92} & 756 \priority{88} & 674.79                   \\
gemma2-it-27b                     & 202 \priority{84} & 73 \priority{83} & 479 \priority{89} & 754 \priority{88} & 175.14                   \\
\rowcolor{gray!20} \textbf{codellama-it-13b}                  & 193 \priority{81} & 77 \priority{88} & 458 \priority{86} & \textbf{728} \priority{85} & \textbf{92.81}                    \\
qwen2-it-72b                      & 202 \priority{84} & 70 \priority{80} & 456 \priority{85} & 728 \priority{85} & 303.35                   \\
codellama-it-34b                  & 192 \priority{80} & 67 \priority{76} & 464 \priority{87} & 723 \priority{84} & 196.09                   \\
phi3-medium-it-14b                & 198 \priority{83} & 77 \priority{88} & 429 \priority{80} & 704 \priority{81} & 106.57                   \\
mistral-nemo-it-2407-12.2b        & 196 \priority{82} & 71 \priority{81} & 433 \priority{81} & 700 \priority{81} & 59.75                    \\
mixtral-v0.1-it-8x7b              & 181 \priority{76} & 71 \priority{81} & 434 \priority{81} & 686 \priority{79} & 302.86                   \\
phi3-small-it-7.3b                & 180 \priority{75} & 66 \priority{75} & 406 \priority{76} & 652 \priority{76} & 81.51                    \\
mistral-v0.3-it-7b                & 177 \priority{74} & 78 \priority{89} & 387 \priority{72} & 642 \priority{75} & 51.51                    \\
llama3.1-it-8b                    & 175 \priority{73} & 69 \priority{78} & 394 \priority{73} & 638 \priority{74} & 170.32                   \\
\rowcolor{gray!20} \textbf{phi3.5-moe-it-41.9b}               & 158 \priority{66} & 68 \priority{77} & 389 \priority{72} & \textbf{615} \priority{71} & 3,574.35                  \\
phi3-mini-it-3.8b                 & 175 \priority{73} & 57 \priority{65} & 372 \priority{69} & 604 \priority{70} & 131.15                   \\
phi3.5-mini-it-3.8b               & 171 \priority{72} & 55 \priority{63} & 344 \priority{64} & 570 \priority{66} & 111.32                   \\
headergen                         & 186 \priority{78} & 56 \priority{64} & 321 \priority{60} & 563 \priority{65} & 18.25                    \\
qwen2-it-7b                       & 167 \priority{70} & 58 \priority{66} & 338 \priority{63} & 563 \priority{65} & 56.61                    \\
codellama-it-7b                   & 164 \priority{69} & 56 \priority{64} & 338 \priority{63} & 558 \priority{64} & 137.01                   \\
hityperdl                         & 163 \priority{68} & 27 \priority{30} & 177 \priority{33} & 367 \priority{43} & 268.4                    \\
gemma2-it-9b   & 22 \priority{9} & 16 \priority{18} & 72 \priority{13} & 110 \priority{13} & 1,521.89                  \\
\rowcolor{gray!20} \textbf{tinyllama-1.1b} & 42 \priority{18} & 7 \priority{8} & 53 \priority{10} & \textbf{102} \priority{12} & 416.63                   \\
gemma2-it-2b   & 21 \priority{9} & 10 \priority{11} & 49 \priority{9} & 80 \priority{9} & 680.03                   
		\\
		\bottomrule              	                                
	\end{tabular}
\end{table}

Table~\ref{tab:type_inference_micro} shows the results of LLMs, HeaderGen, and HiTyper considering the exact-match performance on the \typeevalpy micro-benchmark.
Note that the hybrid analysis tool HiTyper is configured with Type4Py~\citep{mirType4PyPracticalDeep2022c}.
The results highlight that LLMs, particularly recent and larger models, significantly outperform previous approaches like \headergen and HiTyper.
Among the models evaluated, OpenAI's gpt-4o emerges as the best-performing model, correctly inferring 806 of the total 860 type annotations.
This aligns with expectations, as gpt-4o is known for its extensive parameter count and advanced capabilities.
However, its performance comes at the potential cost of speed and computational expense, factors crucial for practical deployment in real-world applications.

Notably, the mistral-large-it-2407-123b model closely follows gpt-4o, correctly predicting 804 type annotations, showing how large open-source models are closing the performance gap with proprietary LLMs. 
This is significant because it implies that with proper tuning and architecture, open-source models can rival closed-source models, providing a potentially more accessible and cost-effective alternative for type inference tasks.
Furthermore, specialized models like CodeLLaMA, particularly the 13B-instruct variant, shows good performance with 728 exact matches, suggesting that fine-tuning models specifically for code-related tasks offers a distinct advantage over general-purpose LLMs like vanilla LLaMA.
In contrast, smaller models such as TinyLlama (1.1B parameters) exhibit poor performance, correctly predicting only 102 annotations, implying that model size is a critical factor for complex tasks like type inference.

From the inference speed perspective, there is a noticeable trade-off between model size, accuracy, and efficiency.
For instance, while larger models like phi3.5-moe-it-41.9b achieve relatively high accuracy, they incur significant inference times (3,574.35 seconds).
In contrast, mid-sized models such as Codellama-it-13b strike a better balance, delivering decent performance with 728 exact matches in a considerably shorter time frame (92.81 seconds).
This suggests that when selecting models for type inference in practice, one must consider not only accuracy but also the computational resources and speed required, especially for large-scale projects or environments with limited hardware.

\subsubsection{TypeEvalPy Autogen Benchmark}

\begin{table}[]
\caption{
Exact match comparison of LLMs for \textbf{type inference} on \typeevalpy autogen benchmark
}

\label{tab:type_inference_autogen}
	\setlength{\tabcolsep}{3.5pt} % Adjusts the space between columns
	\renewcommand{\arraystretch}{1.2}
 \centering 
	
	\begin{tabular}{lrrrrr}
  \multicolumn{6}{r}{\priority{100} 80-100\%,
\priority{70} 60-80\%,
\priority{50} 40-60\%,
\priority{30} 20-40\%,
\priority{10} 0-20\%}  \\  
		\multicolumn{6}{r}{ \textbf{FRT:} Function return type, \textbf{FPT:} Function parameter type, \textbf{LVT:} Local variable type}   \\
		\toprule
		\multicolumn{1}{c}{\textbf{Model}} & \multicolumn{1}{c}{\textbf{FRT}} & \multicolumn{1}{c}{\textbf{FPT}} & \multicolumn{1}{c}{\textbf{LVT}} & \multicolumn{1}{c}{\textbf{Total}} & \multirow{2}{*}{\textbf{Time (s)}}\\
\cmidrule(lr){2-2} \cmidrule(lr){3-3} \cmidrule(lr){4-4} \cmidrule(lr){5-5} 
		\multicolumn{1}{c}{Total} & \multicolumn{1}{c}{17,998} & \multicolumn{1}{c}{896} & \multicolumn{1}{c}{58,374} & \multicolumn{1}{c}{77,268} & \\       
		\midrule
\rowcolor{gray!20} \textbf{mistral-large-it-2407-123b}        & \textbf{16,701} \priority{93}  & \textbf{728} \priority{81}  & \textbf{57,550} \priority{98}  & \textbf{74,979} \priority{97}  & \textbf{28,435}                \\
\rowcolor{gray!20} \textbf{gpt-4o}                            & 16,804 \priority{93}  & 737 \priority{82}  & 56,716 \priority{97}  & 74,257 \priority{96}  & n/a                      \\
\rowcolor{gray!20} \textbf{mixtral-v0.1-it-8x22b}             & 16,424 \priority{91}  & 514 \priority{57}  & 56,550 \priority{97}  & 73,488 \priority{95}  & 22,959                \\
qwen2-it-72b                      & 16,488 \priority{92}  & 629 \priority{70}  & 55,160 \priority{94}  & 72,277 \priority{93}  & 15,877               \\
llama3.1-it-70b                   & 16,648 \priority{92}  & 580 \priority{64}  & 54,445 \priority{93}  & 71,673 \priority{93}  & 14,945               \\
gpt-4o-mini                       & 16,628 \priority{92}  & 643 \priority{71}  & 50,162 \priority{86}  & 67,433 \priority{87}  & n/a                      \\
gemma2-it-27b                     & 16,342 \priority{91}  & 599 \priority{67}  & 49,772 \priority{85}  & 66,713 \priority{86}  & 9,121                \\
\rowcolor{gray!20} \textbf{codestral-v0.1-22b}                & 16,456 \priority{91}  & 706 \priority{79}  & 49,379 \priority{85}  & 66,541 \priority{86}  & 7,437                 \\
codellama-it-34b                  & 15,960 \priority{89}  & 473 \priority{53}  & 48,957 \priority{84}  & 65,390 \priority{84}  & 10,983                \\
mistral-nemo-it-2407-12.2b        & 16,221 \priority{90}  & 526 \priority{59}  & 48,439 \priority{83}  & 65,186 \priority{84}  & 3,227                 \\
mistral-v0.3-it-7b                & 16,686 \priority{93}  & 472 \priority{53}  & 47,935 \priority{82}  & 65,093 \priority{84}  & 2,589                 \\
phi3-medium-it-14b                & 16,802 \priority{93}  & 467 \priority{52}  & 45,121 \priority{77}  & 62,390 \priority{81}  & 6,038                \\
llama3.1-it-8b                    & 16,125 \priority{90}  & 492 \priority{55}  & 44,313 \priority{76}  & 60,930 \priority{79}  & 3,168                 \\
codellama-it-13b                  & 16,214 \priority{90}  & 479 \priority{53}  & 43,021 \priority{74}  & 59,714 \priority{77}  & 4,485                 \\
phi3-small-it-7.3b                & 16,155 \priority{90}  & 422 \priority{47}  & 38,093 \priority{65}  & 54,670 \priority{71}  & 4,400                 \\
qwen2-it-7b                       & 15,684 \priority{87}  & 313 \priority{35}  & 38,109 \priority{65}  & 54,106 \priority{70}  & 4,483                 \\
headergen                         & 14,086 \priority{78}  & 346 \priority{39}  & 36,370 \priority{62}  & 50,802 \priority{66}  & 114                  \\
phi3-mini-it-3.8b                 & 15,908 \priority{88}  & 320 \priority{36}  & 30,341 \priority{52}  & 46,569 \priority{60}  & 3,506                 \\
phi3.5-mini-it-3.8b               & 15,763 \priority{88}  & 362 \priority{40}  & 28,694 \priority{49}  & 44,819 \priority{58}  & 3,631                 \\
codellama-it-7b                   & 13,779 \priority{76}  & 318 \priority{35}  & 29,346 \priority{50}  & 43,443 \priority{56}  & 5,528                 \\
hityperdl                         & 15,765 \priority{88}  & 61 \priority{6}    & 5,365 \priority{9}    & 21,191 \priority{27}  & 6,564                 \\
 gemma2-it-9b   & 1,611 \priority{9}   & 66 \priority{7}    & 5,464 \priority{9}    & 7,141 \priority{9}    & 57,828                \\
 tinyllama-1.1b & 1,514 \priority{8}   & 28 \priority{3}    & 2,699 \priority{5}    & 4,241 \priority{5}    & 13,819                \\
\rowcolor{gray!20}\textbf{mixtral-v0.1-it-8x7b} & 3,235 \priority{18}  & 33 \priority{4}    & 377 \priority{1}    & 3,645 \priority{4}    & 78,215                \\
\rowcolor{gray!20} \textbf{phi3.5-moe-it-41.9b} & 3,090 \priority{17}  & 25 \priority{3}    & 273 \priority{0}    & 3,388 \priority{4}    & 121,617               \\
 gemma2-it-2b   & 1,497 \priority{8}   & 41 \priority{5}    & 1,848 \priority{3}    & 3,386 \priority{4}    & 23,637               
		\\
		\bottomrule                    	                                
	\end{tabular}

\end{table}

\begin{table}[htbp]
    \centering
    \caption{Exact matches comparison between micro-benchmark and autogen-benchmark percentages}
    \renewcommand{\arraystretch}{1.2}
    \begin{tabular}{lrrr}
      \multicolumn{4}{r}{\priority{100} 80-100\%, \priority{70} 60-80\%, \priority{50} 40-60\%, \priority{30} 20-40\%, \priority{10} 0-20\%}  \\  
        \toprule
        \textbf{Model} & \textbf{Micro (\%)} & \textbf{Autogen (\%)} & \textbf{Diff. (\%)} \\
        \midrule
        \multicolumn{4}{c}{\textbf{Models with Consistent Performance} (5\% Delta)} \\
        \midrule
            \rowcolor{gray!20} \textbf{gpt-4o}                            & 93.72 \priority{93} & 96.10 \priority{96} & \textbf{2.38} \\
            \rowcolor{gray!20} \textbf{mistral-large-it-2407-123b}        & 93.49 \priority{93} & 96.97 \priority{96} & \textbf{3.48} \\
            \rowcolor{gray!20} \textbf{gpt-4o-mini}                       & 91.05 \priority{91} & 87.28 \priority{87} & \textbf{-3.77} \\
            llama3.1-it-70b                   & 89.53 \priority{89} & 92.75 \priority{92} & 3.22 \\
            \rowcolor{gray!20} \textbf{codestral-v0.1-22b}                & 88.37 \priority{88} & 86.06 \priority{86} & \textbf{-2.31} \\
            gemma2-it-27b                     & 87.67 \priority{87} & 86.28 \priority{86} & -1.39 \\
            codellama-it-34b                  & 84.07 \priority{84} & 84.64 \priority{84} & 0.57 \\
            phi3-medium-it-14b                & 81.86 \priority{81} & 80.74 \priority{81} & -1.12 \\
            mistral-nemo-it-2407-12.2b        & 81.40 \priority{81} & 84.38 \priority{84} & 2.98 \\
            llama3.1-it-8b                    & 74.19 \priority{74} & 78.89 \priority{78} & 4.70 \\
            qwen2-it-7b                       & 65.47 \priority{65} & 70.02 \priority{70} & 4.55 \\
            \rowcolor{gray!20} \textbf{\headergen}     & 65.47 \priority{65} & 65.73 \priority{65} & \textbf{0.26} \\
            gemma2-it-9b                      & 12.79 \priority{12} & 9.24 \priority{9} & -3.55 \\
        \midrule
        \multicolumn{4}{c}{\textbf{Models that Improved} (Minimum 5\% Increase)} \\
        \midrule
            mixtral-v0.1-it-8x22b             & 87.91 \priority{87} & 95.09 \priority{95} & 7.18 \\
            qwen2-it-72b                      & 84.65 \priority{84} & 93.54 \priority{93} & 8.89 \\
            mistral-v0.3-it-7b                & 74.65 \priority{74} & 84.26 \priority{84} & 9.61 \\
        \midrule
        \multicolumn{4}{c}{\textbf{Models that Deteriorated} (Minimum 5\% Decrease)} \\
        \midrule
            codellama-it-13b                  & 84.65 \priority{84} & 77.26 \priority{77} & -7.39 \\
            \rowcolor{gray!20} \textbf{mixtral-v0.1-it-8x7b}              & 79.77 \priority{79} & 4.72 \priority{4} & \textbf{-75.05} \\
            phi3-small-it-7.3b                & 75.81 \priority{75} & 70.76 \priority{70} & -5.05 \\
            \rowcolor{gray!20} \textbf{phi3.5-moe-it-41.9b}               & 71.51 \priority{71} & 4.38 \priority{4} & \textbf{-67.13} \\
            phi3-mini-it-3.8b                 & 70.23 \priority{70} & 60.27 \priority{60} & -9.96 \\
            phi3.5-mini-it-3.8b               & 66.28 \priority{66} & 57.99 \priority{57} & -8.29 \\
            codellama-it-7b                   & 64.88 \priority{64} & 56.22 \priority{56} & -8.66 \\
            hityperdl                        & 42.67 \priority{42} & 27.43 \priority{27} & -15.24 \\
            tinyllama-1.1b                    & 11.86 \priority{11} & 5.49 \priority{5} & -6.37 \\
        \bottomrule
    \end{tabular}
    \label{tab:exact_match_comparison_difference}
\end{table}

In Table \ref{tab:type_inference_autogen}, we present the results of the same models on the significantly larger and extended \typeevalpy autogen benchmark.
Additionally, in Table \ref{tab:exact_match_comparison_difference}, we list the differences in performance based on the total exact matches between the \typeevalpy micro and autogen benchmarks.

\textbf{Models with Consistent Performance.}
Models in this category demonstrate a maximum delta of 5\% between the micro-benchmark and autogen-benchmark scores.
Notably, gpt-4o and mistral-large-it-2407-123b maintained high exact match across both benchmarks, with deltas of 2.38\% and 3.48\%, respectively. 
The close alignment of these results suggests these models are robust across different testing scenarios, crucial for real-world applications, where model performance needs to generalize across varied datasets.
gpt-4o-mini and codestral-v0.1-22b showed a slight decline of 3.77\% and 2.31\% respectively.
However, these models remained within the acceptable variance threshold, suggesting they are still usable for the type inference tasks.
Additionally, \headergen, with a delta of just 0.26\%, demonstrates the robustness of static analysis tools.

\textbf{Models that Improved.}
Three models showed improvements of more than 5\% in exact matches from the micro-benchmark to the autogen-benchmark.
mixtral-v0.1-it-8x22b improved by 7.18\%, and qwen2-it-72b and mistral-v0.3-it-7b increased by 8.89\% and 9.61\%, respectively.

\textbf{Models that Deteriorated.}
Conversely, several models showed significant performance declines between the two benchmarks.
mixtral-v0.1-it-8x7b and phi3.5-moe-it-41.9b exhibited the largest declines, with mixtral-v0.1-it-8x7b deteriorating by a -75.05\% and phi3.5-moe-it-41.9b by -67.13\%.
The decline in performance indicates a possible overfitting to the \textit{string} datatype that is primarily found in the micro-benchmark.

\begin{tcolorbox}
LLMs, particularly larger models like gpt-4o and \texttt{mistral-large}, demonstrated superior performance in Python type inference compared to traditional static analysis tools such as HeaderGen and HiTyper, on the TypeEvalPy micro-benchmark.
These top-performing LLMs also showed strong consistency when evaluated on the larger TypeEvalPy Autogen benchmark.
\end{tcolorbox}

\section{Discussion}
\label{sec:discussion}

In this section, we discuss the implications of the empirical results observed in the study.
We first analyze call-graph construction in Python and JavaScript, highlighting strengths and weaknesses in different scenarios. 
We then discuss type inference performance in Python, comparing LLMs with traditional tools.
Subsequently, we explore differences in LLM performance between type inference and call-graph analysis, followed by an examination of cross-language disparities, general discussions, and propose avenues for future research.

\subsection{Call Graph Construction in Python: LLMs vs Static Analysis Tools}

In Python, the static analysis tool \pycg consistently outperformed LLMs in constructing call-graphs, with \texttt{mistral-large-it-2407-123b} (mistral-large) ranking highest among the evaluated LLMs. Table \ref{tab:comparison_pycg_mistral-large-it-2407-123b} compares \pycg and \texttt{mistral-large} across selected categories from the \pycg micro-benchmark, chosen specifically for similarities and differences in tool performance. 
Furthermore, in Table \ref{tab:failure_patterns_python}, we list specific patterns in which LLMs struggled.

Within the \textit{returns} category, both \pycg and mistral-large accurately resolved cases involving direct function returns and imported functions. However, \texttt{mistral-large} failed to handle scenarios with indirect imports through intermediate modules correctly, introducing false positives and omissions, while \pycg resolved these cases correctly. 
In complex multi-level return structures, as illustrated by the \textit{return\_complex} test case, \texttt{mistral-large} missed call edges, resulting in unsoundness, whereas \pycg successfully identified all call relationships.

Complex return constructs, particularly those involving Python's generator feature using yield statements, are common in real-world projects.
Yield-based generator returns are the third most frequent functional feature in the dataset consisting of over 3.1 million Python files from 51,493 GitHub repositories created by \cite{yang_complex_2022}.
This highlights the real-world significance of accurately handling complex return constructs.
The \textit{dicts} category demonstrated stronger performance by \texttt{mistral-large}, which achieved perfect completeness, soundness, and exact match rates, surpassing \pycg. Particularly notable was \texttt{mistral-large}'s correct handling of dictionary updates using the \textit{update()} method, a scenario where \pycg incorrectly missed a call edge. 

Python dictionary features contribute to efficient and flexible data storage. Beyond direct method calls like \texttt{update()}, dictionary comprehensions serve as an efficient way to create and manipulate dictionaries in real-world code. 
The study by \cite{yang_complex_2022} categorizes dictionary comprehension constructs as functional features and recorded 81,763 occurrences in their dataset, ranking them as the fifth most frequent functional feature.

In contrast, within the \textit{functions} category, both \pycg and \texttt{mistral-large} demonstrated perfect accuracy, indicating comparable performance in resolving direct calls, variable assignments, and cross-module function imports.

\begin{tcolorbox}
In constructing call-graphs for Python, \pycg consistently outperformed LLMs overall, with \texttt{mistral-large} showing competitive performance in specific cases such as dictionary updates, although it exhibited unsoundness in complex return structures and indirect imports where \pycg remained accurate.
\end{tcolorbox}

\begin{table}[]
    \centering
    \renewcommand{\arraystretch}{1.2}
    \setlength{\tabcolsep}{4pt} % Adjusts the space between columns
    \caption{Category-wise call-graph construction performance on Python micro-benchmark}
    \label{tab:comparison_pycg_mistral-large-it-2407-123b}
    \begin{tabular}{l|rrr|rrr}
    \multicolumn{7}{r}{\priority{100} 80-100\%, \priority{70} 60-80\%, \priority{50} 40-60\%, \priority{30} 20-40\%, \priority{10} 0-20\%} \\  
    \toprule
    \multirow{2}{*}{\textbf{Category}} & \multicolumn{3}{c|}{\textbf{PyCG}} & \multicolumn{3}{c}{\textbf{mistral-large-it-2407-123b}} \\
    \cmidrule(lr){2-4} \cmidrule(lr){5-7}
    & \textbf{Complete} & \textbf{Sound} & \textbf{E.M.} & \textbf{Complete} & \textbf{Sound} & \textbf{E.M.} \\
    \midrule
    \textbf{returns}          & 4/4 \priority{100}  & 4/4 \priority{100}  & 24/24 \priority{100} & 3/4 \priority{77}  & 2/4 \priority{50} & 22/24 \priority{92} \\
    \textbf{dicts}  & 9/12 \priority{75}  & 11/12 \priority{92}  & 40/41 \priority{98}  & 12/12 \priority{100}  & 12/12 \priority{100}  & 41/41 \priority{100}  \\
    \textbf{functions}          & 4/4 \priority{100} & 4/4 \priority{100}  & 9/9 \priority{100} & 4/4 \priority{100}  & 4/4 \priority{100} & 9/9 \priority{100} \\
    \bottomrule
    \end{tabular}
\end{table}

\begin{table}[]
\centering
\caption{Challenging patterns that affect tool performance on Python benchmark}
\label{tab:failure_patterns_python}
\renewcommand{\arraystretch}{1.15}
\setlength{\tabcolsep}{8pt}
\begin{tabular}{@{} l >{\raggedright\arraybackslash}p{8.5cm} @{}}
\toprule
\textbf{Sub-category} & \textbf{Concise example} \\ \midrule

\multicolumn{2}{c}{\textbf{Returns}} \\ \midrule

\texttt{return\_complex} &
\begin{minipage}[c]{\linewidth}
\begin{lstlisting}[language=Python,numbers=none,frame=none,basicstyle=\footnotesize\ttfamily]
def func3():
    return func2

def func4(a):
    return func3()

...

func4()()
\end{lstlisting}
\end{minipage} \\ \addlinespace[4pt]

\midrule
\multicolumn{2}{c}{\textbf{Dicts}} \\ \midrule

\texttt{update} &
\begin{minipage}[c]{\linewidth}
\begin{lstlisting}[language=Python,numbers=none,frame=none,basicstyle=\footnotesize\ttfamily]
def func1():
    pass

def func2():
    pass

d = {"a": func1}

d.update({"a": func2})
d["a"]()
\end{lstlisting}
\end{minipage} \\ \addlinespace[4pt]

% \midrule
% \multicolumn{2}{c}{\textbf{Functions}} \\ \midrule

% \texttt{imported\_call} &
% \begin{minipage}[c]{\linewidth}
% \begin{lstlisting}[language=Python,numbers=none,frame=none,basicstyle=\footnotesize\ttfamily]

% \end{lstlisting}
% \end{minipage} \\ \addlinespace[4pt]

\bottomrule
\end{tabular}
\end{table}

\subsection{Call Graph Construction in JavaScript: LLMs vs Static Analysis Tools}

In JavaScript, the static analysis tool Jelly outperformed LLMs in terms of soundness and exact match rates, whereas the TAJS static analysis tool produced poor results due to its lack of support for modern JavaScript features and inactivity in recent years. Table \ref{tab:comparison_jelly_mistral-large-it-2407-123b} compares Jelly and \texttt{mistral-large} on the \swarmjs micro-benchmark across categories chosen for their distinct and overlapping tool behaviours.

Table \ref{tab:challenging_patterns_javascript} lists specific challenging patterns where LLMs failed.
In the \textit{arguments} category, Jelly achieved perfect soundness, identifying all valid call edges, and completeness in 4 out of 10 test cases, while \texttt{mistral-large} matched Jelly's completeness but was sound in only half of the cases. Both tools effectively handled direct function passing and default arguments, though \texttt{mistral-large} struggled with indirect argument flows and cross-file imports.

Modern JavaScript features related to argument handling, such as default parameters and spread arguments, are widely adopted, appearing in over 56\% and 60\% of projects, respectively, in a study of 158 open-source systems \citep{lucas_understanding_2025}.
Arrow Function Declaration, which offer concise argument syntax, is a highly popular feature, present in nearly 88\% of projects in this dataset.
The prevalence of these features underscores the importance of correctly resolving call graphs involving diverse argument patterns.

The \textit{classes} category showed Jelly's clear superiority, achieving soundness in all test cases, whereas \texttt{mistral-large} showed significant challenges, particularly with inheritance, chained attribute references, and destructured assignments. In tests involving inheritance and method assignment, such as \textit{base\_class\_calls\_child}, Jelly reliably resolved call edges, contrasting \texttt{mistral-large}'s frequent misses.
The adoption of class syntax in JavaScript is substantial. A study by \cite{nishiura_analyzing_2024} on 636 GitHub projects found that over half of the projects use class syntax, indicating a shift towards class-based programming.
Furthermore, class inheritance (extends) is widely utilized, appearing in nearly 70\% of projects that use classes.

Within the \textit{objects} category, both Jelly and \texttt{mistral-large} effectively managed direct object access and calls via parameter returns. However, \texttt{mistral-large} surpassed Jelly in cases involving dynamically derived object keys from function parameters or external modules. Nevertheless, Jelly correctly handled a test case involving type coercion, which \texttt{mistral-large} did not, thus missing a call edge.
A study by \cite{lucas_understanding_2025} found that object features like object destructuring are common, appearing in nearly 69\% of projects in a dataset of 158 open-source JavaScript projects.

\begin{tcolorbox}
In JavaScript call-graph construction, Jelly consistently outperformed LLMs, including \texttt{mistral-large}, particularly in classes and complex argument flows, while \texttt{mistral-large} demonstrated competitive handling of dynamic object keys but showed significant unsoundness in inheritance and indirect data flows.
\end{tcolorbox}

\begin{table}[]
    \centering
    \renewcommand{\arraystretch}{1.2}
    \setlength{\tabcolsep}{4pt} % Adjusts the space between columns
    \caption{Category-wise call-graph construction performance on JavaScript micro-benchmark}
    \label{tab:comparison_jelly_mistral-large-it-2407-123b}
    \begin{tabular}{l|rrr|rrr}
    \multicolumn{7}{r}{\priority{100} 80-100\%, \priority{70} 60-80\%, \priority{50} 40-60\%, \priority{30} 20-40\%, \priority{10} 0-20\%} \\  
    \toprule
    \multirow{2}{*}{\textbf{Category}} & \multicolumn{3}{c|}{\textbf{Jelly}} & \multicolumn{3}{c}{\textbf{mistral-large-it-2407-123b}} \\
    \cmidrule(lr){2-4} \cmidrule(lr){5-7}
    & \textbf{Complete} & \textbf{Sound} & \textbf{E.M.} & \textbf{Complete} & \textbf{Sound} & \textbf{E.M.} \\
    \midrule
    \textbf{args}  & 4/10 \priority{40}  & 10/10 \priority{100}  & 37/37 \priority{100}  & 4/10 \priority{40}  & 5/10 \priority{50}  & 30/37 \priority{81}  \\
    \textbf{classes}          & 5/21 \priority{24} & 20/21 \priority{95}  & 87/92 \priority{94} & 1/21 \priority{5}  & 3/21 \priority{14} & 55/92 \priority{60} \\
    \textbf{objects}          & 7/12 \priority{58}  & 9/12 \priority{75}  & 37/41 \priority{90} & 11/12 \priority{92}  & 11/12 \priority{92} & 40/41 \priority{97} \\
    \bottomrule
    \end{tabular}
\end{table}

\begin{table}[htbp]
\centering
\caption{Challenging patterns impacting call-graph construction on JavaScript benchmark}
\label{tab:challenging_patterns_javascript}
\renewcommand{\arraystretch}{1.15}
\setlength{\tabcolsep}{8pt}
\begin{tabular}{@{} l >{\raggedright\arraybackslash}p{9cm} @{}}
\toprule
\textbf{Sub-Category} & \textbf{Example Snippet} \\ 
\midrule
\multicolumn{2}{c}{\textbf{Arguments}} \\ 
\midrule

\texttt{nested\_call} &
\begin{minipage}[c]{\linewidth}
\begin{lstlisting}[language=JavaScript,numbers=none,frame=none,basicstyle=\footnotesize\ttfamily]
function paramFunc(a) {
    a();
}
function func(a) {
    a(function nestedFunc() {...});
}
const b = paramFunc;
const c = func;
c(b);
\end{lstlisting}
\end{minipage} \\ \addlinespace[4pt]
\midrule
\texttt{imported\_call} &
\begin{minipage}[c]{\linewidth}
\begin{lstlisting}[language=JavaScript,numbers=none,frame=none,basicstyle=\footnotesize\ttfamily]
import { func } from "./to_import.js";

function paramFunc() { }
func(paramFunc);
\end{lstlisting}
\end{minipage} \\ \addlinespace[4pt]

\midrule
\multicolumn{2}{c}{\textbf{Classes}} \\
\midrule

\texttt{base\_class\_calls\_child} &
\begin{minipage}[c]{\linewidth}
\begin{lstlisting}[language=JavaScript,numbers=none,frame=none,basicstyle=\footnotesize\ttfamily]
class A {
    func() {
        if (this.child) { this.child(); }
    }
}
class B extends A {
    constructor() {
        super();
        this.child = this.func2;
    }
    func2() {...}
}
...
  
const b = new B();
b.func();
\end{lstlisting}
\end{minipage} \\ \addlinespace[4pt]

% More examples from Classes here

\midrule
\multicolumn{2}{c}{\textbf{Objects}} \\
\midrule

\texttt{type\_coercion} &
\begin{minipage}[c]{\linewidth}
\begin{lstlisting}[language=JavaScript,numbers=none,frame=none,basicstyle=\footnotesize\ttfamily]
function func1() {...}
function func2() {...}

let d = {1: func1, "1": func2};
d[1](); 
\end{lstlisting}
\end{minipage} \\ \addlinespace[4pt]

% More examples from Objects here if needed

\bottomrule
\end{tabular}
\end{table}

\subsection{Type Inference in Python: LLMs vs Static Analysis Tools}
\label{sec:discussion_type_inference}

Table \ref{tab:type_inference_micro} lists the type-inference performance of \texttt{mistral-large} and \headergen{} on Micro and Autogen benchmarks.
The analysis concentrates on three language constructs: assignments, decorators, and generators, which show a large performance difference.

\headergen{}’s errors arise primarily from the absence of modelling edge-case language constructs.
Table \ref{tab:failure_patterns} shows complex cases where \headergen failed to infer the types correctly.
In assignments, it lacks rules for augmented updates, star unpacking, and tuples that are repeatedly unpacked and repacked, where \headergen falls back to the generic type \texttt{Any}.
In decorators, \headergen misses the decorators that change a function’s signature or return type.
In generators, it does not track the return type of a user-defined \texttt{\_\_next\_\_} method to the type produced by the function that is yielding.

Developing and maintaining such fine-grained modelling of language constructs is laborious.
Static analysers, therefore, default to conservative approximation as \texttt{Any}.
In contrast, an LLM acquires these behaviors implicitly through large-scale exposure to real-world repositories, learning that \texttt{int += int} remains an \texttt{int} and that a wrapper can replace a function’s return type, it therefore preserves precise types where \headergen{} widens to \texttt{Any}.

\begin{tcolorbox}
In Python type inference, \texttt{mistral-large} surpassed the static analysis tool \headergen{} by accurately preserving precise types in complex constructs such as augmented assignments, decorators, and generators, whereas \headergen{} defaulted to conservative approximations due to limited modelling of edge-case language constructs.
\end{tcolorbox}

\begin{table}[htbp]
\centering
\caption{Category-wise type-inference performance across Micro and Autogen benchmarks}
\label{tab:type_inference_micro}
\setlength{\tabcolsep}{1.8pt}
\renewcommand{\arraystretch}{1.2}
\begin{tabular}{lrrr|rrr|rrr}
\multicolumn{10}{r}{\priority{100} 80--100,
\priority{70} 60--80,
\priority{50} 40--60,
\priority{30} 20--40,
\priority{10} 0--20} \\[-2pt]
\multicolumn{10}{r}{\textbf{FRT:} Function-return type, \textbf{FPT:} Function-parameter type, \textbf{LVT:} Local-variable type}\\
\toprule
 & \multicolumn{3}{c}{\textbf{Ground Truth}} & \multicolumn{3}{c}{\textbf{mistral-large-it-2407-123b}} & \multicolumn{3}{c}{\textbf{\headergen{}}}\\
\cmidrule(lr){2-4}\cmidrule(lr){5-7}\cmidrule(lr){8-10}
\textbf{Category} & FRT & FPT & LVT & FRT (\%) & FPT (\%) & LVT (\%) & FRT (\%) & FPT (\%) & LVT (\%)\\
\midrule
\multicolumn{10}{c}{\textbf{Micro-benchmark}}\\
\midrule
\textbf{assignments}
& 22 & 4  & 56
& 95.5 \priority{100} & 100.0 \priority{100} & 100.0 \priority{100}
& 68.2 \priority{70}  & 25.0 \priority{30}  & 58.9 \priority{50}\\
\textbf{decorators}
& 29 & 17 & 12
& 86.2 \priority{100} & 70.6 \priority{70} & 58.3 \priority{50}
& 37.9 \priority{30} & 35.3 \priority{30} & 16.7 \priority{10}\\
\textbf{generators}
& 13 & 8  & 49
& 84.6 \priority{100} & 100.0 \priority{100} & 98.0 \priority{100}
& 69.2 \priority{70}  & 50.0 \priority{50}  & 34.7 \priority{30}\\
\midrule
\multicolumn{10}{c}{\textbf{Autogen Benchmark}}\\
\midrule
\textbf{assignments}
& 8,308 & 8 & 25,357
& 91.3 \priority{100} & 100.0 \priority{100} & 99.9 \priority{100}
& 77.6 \priority{70}  & 50.0 \priority{50}  & 62.6 \priority{70}\\
\textbf{decorators}
& 748 & 269 & 494
& 77.5 \priority{70}  & 82.9 \priority{100} & 75.7 \priority{70}
& 36.1 \priority{30}  & 25.7 \priority{30} & 6.1  \priority{10}\\
\textbf{generators}
& 49  & 24 & 186
& 89.8 \priority{100} & 100.0 \priority{100} & 91.9 \priority{100}
& 79.6 \priority{70} & 79.2 \priority{70} & 34.4 \priority{30}\\
\bottomrule
\end{tabular}
\end{table}

\begin{table}[htbp]
\centering
\caption{Challenging patterns that trigger \headergen{} failures}
\label{tab:failure_patterns}
\renewcommand{\arraystretch}{1.15}
\setlength{\tabcolsep}{8pt}
\begin{tabular}{@{} l >{\raggedright\arraybackslash}p{7.8cm} @{}}
\toprule
\textbf{Sub-category} & \textbf{Concise example} \\ \midrule
\multicolumn{2}{c}{\textbf{Assignments}}\\
\midrule
\texttt{augmented} &
\begin{minipage}[c]{\linewidth}
\begin{lstlisting}[language=Python,numbers=none,frame=none,basicstyle=\footnotesize\ttfamily]
a += 3
\end{lstlisting}
\end{minipage} \\ \addlinespace[4pt]
\midrule

\texttt{starred} &
\begin{minipage}[c]{\linewidth}
\begin{lstlisting}[language=Python,numbers=none,frame=none,basicstyle=\footnotesize\ttfamily]
a, *b, c = f1, f2, f3
\end{lstlisting}
\end{minipage} \\ \addlinespace[4pt]
\midrule

\texttt{nested\_unpack} &
\begin{minipage}[c]{\linewidth}
\begin{lstlisting}[language=Python,numbers=none,frame=none,basicstyle=\footnotesize\ttfamily]
(x, (y, z)) = (1, (2, 3))
\end{lstlisting}
\end{minipage} \\ \addlinespace[4pt]
\midrule

\texttt{recursive\_tuple} &
\begin{minipage}[c]{\linewidth}
\begin{lstlisting}[language=Python,numbers=none,frame=none,basicstyle=\footnotesize\ttfamily]
t1 = (1, 2); (a, b) = t1
\end{lstlisting}
\end{minipage} \\ \addlinespace[6pt]
\midrule

\multicolumn{2}{c}{\textbf{Decorators}}\\
\midrule

\texttt{classes} &
\begin{minipage}[c]{\linewidth}
\begin{lstlisting}[language=Python,numbers=none,frame=none,basicstyle=\footnotesize\ttfamily]
@decorator
class C: ...
\end{lstlisting}
\end{minipage} \\ \addlinespace[4pt]
\midrule

\texttt{nested\_decorators} &
\begin{minipage}[c]{\linewidth}
\begin{lstlisting}[language=Python,numbers=none,frame=none,basicstyle=\footnotesize\ttfamily]
@d1
@d2
def g(): ...
\end{lstlisting}
\end{minipage} \\ \addlinespace[6pt]
\midrule

\multicolumn{2}{c}{\textbf{Generators}}\\
\midrule

\texttt{yield\_function} &
\begin{minipage}[c]{\linewidth}
\begin{lstlisting}[language=Python,numbers=none,frame=none,basicstyle=\footnotesize\ttfamily]
def f(): return 

def g(n):                  
    for _ in range(n):
        yield 5

for fn in g(10):
    print(fn())
\end{lstlisting}
\end{minipage} \\ \addlinespace[4pt]
\midrule

\texttt{yield\_next} &
\begin{minipage}[c]{\linewidth}
\begin{lstlisting}[language=Python,numbers=none,frame=none,basicstyle=\footnotesize\ttfamily]
def squares():
    n = 1
    while True:
        yield n * n
        n += 1

gen = squares()
for _ in range(5):
    a = next(gen)
\end{lstlisting}
\end{minipage} \\
\bottomrule
\end{tabular}
\end{table}

\subsection{LLM Performance Differences: Type Inference vs. Call Graph Analysis}
The empirical results demonstrate that LLMs show notably stronger performance in type inference tasks compared to call-graph analysis.
However, explaining this behavior of LLMs is challenging, as their performance often emerges from complex interactions between training data, model architecture, and task formulation. 
Nonetheless, a likely explanation lies in the nature of LLM training: Python type annotations are embedded directly within source code and naturally align with next-token prediction objectives, enabling models to learn type patterns during pretraining. 
Although the micro-benchmark used in this study was newly created, lacked in-code annotations, and was unlikely to have been seen during pretraining, the LLMs were still able to generalize and perform well.
Type inference is also benefited in certain instances, such as local variable assignments, where complex understanding of global program behavior is not always necessary, and types can often be inferred from the nearby context.
By contrast, call-graph construction requires reasoning about control flows and structural relationships, which are harder to infer from token sequences alone, presenting greater challenges for LLMs.
These capabilities are less likely to emerge purely from scale and pretraining on language-like sequences \citep{bertiEmergentAbilitiesLarge2025}.
Additionally, \cite{obrienMeasuringEmergentCapabilities2024} found no emergent improvements for software engineering tasks such as bug fixing or code analysis with increased model scale, suggesting that such tasks demand reasoning mechanisms not easily captured by current LLM architectures or pretraining regimes.

\begin{tcolorbox}
LLMs demonstrated stronger performance in type inference than call-graph analysis, likely due to the alignment of type information with token prediction objectives during pretraining, whereas call-graph construction requires complex global reasoning that is less accessible through local token patterns.
\end{tcolorbox}

\subsection{Cross-language Performance Disparities}
A consistent observation throughout our experiments is that LLM performance is notably better on Python code than on JavaScript.
Early studies have indicated that LLMs perform better on code generation tasks in Python compared to JavaScript~\citep{buscemi_comparative_2023}, likely due to Python’s simpler syntax and more uniform structure.
This has been further speculated by prior work~\citep{chen2021evaluatinglargelanguagemodels}, which highlights the influence of cleaner semantics and stronger conventions in Python.
Despite these insights, a more systematic investigation is required to understand the underlying causes of this performance gap.

\subsection{Implications for Type Inference in Dynamic Languages}
Our findings suggest that LLMs are surpassing traditional tools in type inference tasks, especially in dynamically typed languages like Python.
This has significant implications for large codebases, where manual annotation is often infeasible.
Accurate type inference can substantially improve code readability, enable better tooling (e.g., code completion, static analysis), and facilitate the gradual adoption of type annotations in legacy projects.

Models such as \texttt{gpt-4o} and \texttt{mistral-large-it-2407-123b} demonstrate superior accuracy in inferring types.
This capability suggests a potential shift in how type information is extracted and utilized within development workflows, moving from static, rule-based systems toward data-driven, context-aware assistants.
However, the deployment of these models is not without challenges. Their computational requirements, including high memory usage and inference latency, can make LLMs difficult to integrate into resource-constrained environments such as continuous integration pipelines or lightweight IDE plugins.
Furthermore, concerns around determinism, explainability, and security (particularly with closed-source models) must also be considered when using LLMs in production tooling.

\begin{tcolorbox}
LLMs show significant performance improvements in type inference for dynamic languages like Python, indicating a potential toward data-driven analysis approaches in development workflows, although practical deployment faces challenges related to resource demands, determinism, and security.    
\end{tcolorbox}

\subsection{Trade-offs Between Model Accuracy and Efficiency}
Interestingly, mid-sized models like \texttt{codellama-it-13b} and \texttt{codestral-v0.1-22b} offer a more balanced trade-off, achieving competitive accuracy with lower inference time.
These results imply that specialized fine-tuning and architectural choices can lead to performance levels comparable to, or even better than, general-purpose proprietary models.
Conversely, the notably poor performance of lightweight models like \texttt{tinyllama-1.1b} suggests that there is a lower bound on model complexity necessary for robust type inference.
Results indicate that these smaller models lack the representational capacity needed to capture the complex code patterns that type inference demands, particularly in dynamically typed languages where explicit type hints are sparse. 
This observation suggests that while lightweight models may be attractive for extremely resource-constrained settings, they may not yet be viable replacements when inference precision is critical.
In real-world applications, smaller models with moderate accuracy and faster inference times may be more appropriate in iterative development environments.

\begin{tcolorbox}
Mid-sized models such as \texttt{codellama-it-13b} and \texttt{codestral-v0.1-22b} achieve a balance between accuracy and efficiency, whereas smaller models like \texttt{tinyllama-1.1b} lack the capacity for reliable type inference, indicating that a minimum model complexity is necessary for precision in dynamic languages.
\end{tcolorbox}

\subsection{Scalability and Deployment Considerations}
Most LLMs evaluated in this study have over seven billion parameters, which typically require multi-GPU setups or specialized hardware (e.g., high-memory A100 or H100 nodes) to perform inference at acceptable speeds. 
This makes them impractical for deployment on standard single-GPU machines commonly used by individual developers.
In contrast, traditional tools such as \pycg{} and \headergen{} can be executed efficiently in such environments, making them more viable for integration into developer workflows where hardware resources are limited.
This gap points to the need for either lighter, more optimized LLM variants specifically designed for developer tooling or hybrid approaches that combine traditional static analysis with targeted LLM augmentation only when necessary.

\begin{tcolorbox}
Due to their high hardware demands, current LLMs are impractical for standard single-GPU deployments, highlighting the need for lighter LLM variants or hybrid approaches that combine static analysis with selective LLM augmentation to support resource-constrained developer environments.
\end{tcolorbox}

\subsection{Towards Hybrid Analysis: LLMs as Enhancers of Static Tools}
Given their success in type inference, LLMs could serve as auxiliary tools to enrich traditional static analysis pipelines rather than as replacements.
Accurately inferred types could enhance call-graph construction, especially in cases involving dynamic dispatch or polymorphism.
By integrating inferred type annotations into SA pipelines, one could improve the precision and recall of downstream analyses.
This hybrid approach—combining LLMs' contextual understanding with the rigor of SA tools—presents a promising direction for future work.
However, realizing this vision will require careful system design. 
Issues such as calibration of confidence thresholds for LLM outputs, handling conflicting inferences, and maintaining transparency and auditability within SA workflows must be addressed.

\begin{tcolorbox}
LLMs could supplement rather than replace static analysis tools by providing inferred types to improve downstream analyses like call-graph construction, though achieving effective hybrid integration requires addressing challenges related to confidence calibration and transparency.
\end{tcolorbox}

\section{Threats to Validity}
\label{sec:ttv}
We acknowledge the following limitations and threats to the validity of our study:
\begin{itemize}
    \item We applied the same prompt to all models, which may not have optimized performance for each individual model. Tailored prompts could potentially extract better results from specific models.
    \item Open-source models frequently deviated from the expected output formats provided in the prompt. To mitigate this, we manually identified response patterns and added a preprocessing step to standardize the format. However, this approach may not account for all variations, further underscoring the challenge of consistently generating structured data with LLMs.
    \item While we tested several prompts iteratively, our approach did not focus exclusively on optimizing prompt engineering. A dedicated experiment to explore different prompting strategies could lead to better results. Our modular framework can serve as a foundation for future research aimed at refining prompts to improve performance.
    \item We used greedy search for token prediction, always selecting the highest-probability token. Future research could explore higher temperature settings and incorporate a voting mechanism to identify the best output, potentially yielding better results.
    \item While micro-benchmarks are useful for isolating and evaluating specific aspects of system performance, they may miss the complexity and variability of real-world workloads and use cases.
    Therefore, we extended TypeEvalPy with auto-generation capabilities to improve the 
    type diversity of the micro-benchmark. It's hard to ensure that all variations were
    considered in this study and that the 
    results will generalize, but we made efforts
    to extend considerably the amount of use cases 
    previously available.    
    \item Extending \typeevalpy with auto-generation capabilities required significant human effort to create the initial templates. While two authors reviewed these templates, they may still be subject to human error. However, we believe that any minor mistakes in the templates are unlikely to have a significant impact on the overall results of this study.
\end{itemize}

\section{Conclusion}
\label{sec:conclusion}

This study provides a comprehensive evaluation of LLMs in static analysis tasks, particularly call-graph construction and type inference, using enhanced micro-benchmarks across Python and JavaScript programs.
Our results reaffirm that while, LLMs offer promising capabilities in various software engineering tasks, for Python traditional static analysis methods remain more effective for call-graph construction.
Similar to findings in previous studies, LLMs have yet to surpass the efficiency of static tools like PyCG and Jelly for this task.

Interestingly, our analysis also highlights a notable performance difference between LLMs' handling of Python and JavaScript code, with LLMs generally performing better on Python.
One possible explanation is that LLMs may inherently handle Python more effectively due to the language's widespread use in LLM training datasets.
Yet, further investigation is required to fully understand this performance gap.

In type inference tasks, LLMs demonstrated a clear advantage over traditional tools where models like gpt-4o and mistral-large-it-2407-123b excelled.
However, their large computational demands limit their practicality in resource-constrained environments.
Notably, smaller specialized models like codestral-v0.1-22b showed competitive performance, highlighting the potential for optimization.

This study demonstrates the potential of LLMs in software engineering tasks, while also emphasizing their limitations and the continued strengths of traditional methods.
Future research should explore hybrid approaches that combine the strengths of LLMs and static analysis to further advance the field, for instance, by using type inference capabilities of LLMs with traditional static analysis techniques to improve call-graph construction, especially in handling dynamic dispatch and polymorphism.
\vfil

\appendix
\section{Prompts}
\label{sec:prompts}

\subsection{Type Inference Prompts}

\begin{minipage}{\linewidth}
\begin{lstlisting}[style=promptstyle, caption=Prompt for type inference with JSON output format, label=lst:prompt_json_types]
You will be provided with the following information:
1. Python code. The sample is delimited with triple backticks.
2. Sample JSON containing type inference information for the Python code in a specific format.
3. Examples of Python code and their inferred types. The examples are delimited with triple backticks. These examples are to be used as training data.

Perform the following tasks:
1. Infer the types of various Python elements like function parameters, local variables, and function return types according to the given JSON format with the highest probability.
2. Provide your response in a valid JSON array of objects according to the training sample given. Do not provide any additional information except the JSON object.
3. {format_instructions}

Example Python Code:
```main.py
{code snippet}
```

Example JSON Response:
```
{example response in JSON format}
```
\end{lstlisting}
\end{minipage}

\begin{minipage}{\linewidth}
\begin{lstlisting}[style=promptstyle, caption=Prompt for type inference in question-answer format (part 1 of 2), label=lst:prompt_qa_types_1]
## Task Description

**Objective**: Examine and identify the data types of various elements such as function parameters, local variables, and function return types in the given Python code.

**Instructions**:
1. For each question below, provide a concise, one-word answer indicating the data type.
2. For arguments and variables inside a function, list every data type they take within the current program context as a comma separated list.
3. Do not include additional explanations or commentary in your answers.
\end{lstlisting}
\end{minipage}

\begin{minipage}{\linewidth}
\begin{lstlisting}[style=promptstyle, caption=Prompt for type inference in question-answer format (part 2 of 2), label=lst:prompt_qa_types_2]
**Example Python Code**:
```python
a = 1
b = 1.0
c = "hello"
```

**Example Questions**:
1. What is the type of the variable 'a' at line 1, column 1? Reply in one word.
2. What is the type of the variable 'b' at line 2, column 1? Reply in one word.
3. What is the type of the variable 'c' at line 3, column 1? Reply in one word.

**Example Answers**:
1. int
2. float
3. str

**Python Code Provided**:
{code}

**Questions**:
{questions}

**Format for Answers**:
- Provide your answer next to each question number, using only one word.
- Example:
    1. int
    2. float
    3. str

**Your Answers**:
{answers}
\end{lstlisting}
\end{minipage}

\begin{minipage}{\linewidth}
\begin{lstlisting}[style=promptstyle, caption=Example of an actual full prompt for type inference in question-answer format used in the study, label=lst:actual_prompt]
## Task Description

**Objective**: Examine and identify the data types of various elements such as function parameters, local variables, and function return types in the given Python code.

**Instructions**:
1. For each question below, provide a concise, one-word answer indicating the data type.
2. For arguments and variables inside a function, list every data type they take within the current program context as a comma separated list.
3. Do not include additional explanations or commentary in your answers.

**Example Python Code**:
```python
a = 1
b = 1.0
c = "hello"
```

**Example Questions**:
1. What is the type of the variable 'a' at line 1, column 1? Reply in one word.
2. What is the type of the variable 'b' at line 2, column 1? Reply in one word.
3. What is the type of the variable 'c' at line 3, column 1? Reply in one word.

**Example Answers**:
1. int
2. float
3. str

**Python Code Provided**:
```main.py
def param_func():
    return "Hello from param_func"


def func(a):
    return a()


b = param_func
c = func(b)
```

**Questions**:
1. What is the return type of the function 'param_func' at line 1, column 5?
2. What is the return type of the function 'func' at line 5, column 5?
3. What is the type of the parameter 'a' at line 5, column 10, within the function 'func'?
4. What is the type of the variable 'b' at line 9, column 1?
5. What is the type of the variable 'c' at line 10, column 1?

**Format for Answers**:
- Provide your answer next to each question number, using only one word.
- Example:
    1. int
    2. float
    3. str

**Your Answers**:
1.
2.
3.
4.
5.
\end{lstlisting}
\end{minipage}

\subsection{Call-graph Prompts}

\begin{minipage}{\linewidth}
\begin{lstlisting}[style=promptstyle, caption=Prompt for call-graph task in question-answer format (Part 1 of 2), label=lst:prompt_qa_cg_1]
## Task Description

**Objective**: Examine and identify the function calls in the given {language} code and answer the questions.

**Instructions**:
1. For each question below, provide a concise answer indicating the function calls.
2. List every function call as a comma separated list.
3. Do not include additional explanations or commentary in your answers.
4. Include both explicit and implicit function calls in your answers. An implicit function call is a function that is called as a result of another operation, such as the __init__ method being called when an object is created.
5. If a function is called through an alias or a reference, identify and list the actual function that is called after resolving the alias.
6. If a passed argument is not invoked within the function, do not include the function call in the answer.
7. Example of {language} code, questions, and answers are given below. This example should be used as training data.
\end{lstlisting}
\end{minipage}
\begin{minipage}{\linewidth}
\begin{lstlisting}[style=promptstyle, caption=Prompt for call-graph task in question-answer format (Part 2 of 2), label=lst:prompt_qa_cg_2]
**Format for Answers**:
- Provide your answer next to each question number, using only one word.
- Do not include the questions in your answer.
- Example:
    1. simple.func
    2. simple.examplefunc
    3. func

**Example Questions**:
1. What are the module level function calls in the file "main.py"?
2. What are the function calls inside the "main.return_func" function in the file "main.py"?
3. What are the function calls inside the "main.func" function in the file "main.py"?

**Example Answers**:
1. main.func, main.return_func
2. main.func
3.

**{language} Code Provided**:
{code}

**Questions**:
{questions}

**Answers**:
{answers}
\end{lstlisting}
\end{minipage}

\begin{minipage}{\linewidth}
\begin{lstlisting}[style=promptstyle, caption=Prompt for flow-sensitive call-graph task in question-answer format (Part 1 of 2), label=lst:prompt_qa_cs_1]
## Task Description

**Objective**: Examine and identify the fully qualified names of function calls in the given {language} code, including class methods with both the class name and the method name.

**Instructions**:
1. For each question below, provide a concise answer indicating the fully qualified name of function call in that line, including class name for methods.
2. List every function call in that line as a comma separated list.
3. Do not include additional explanations or commentary in your answers.
4. Example of {language} code, questions, and answers are given below. This example should be used as training data.
\end{lstlisting}

\begin{lstlisting}[style=promptstyle, caption=Prompt for flow-sensitive call-graph task in question-answer format (Part 2 of 2), label=lst:prompt_qa_cs_2]
**Format for Answers**:
- Provide your answer next to each question number, using only one word.
- Add the line number of where the function is defined in the program to the answer separated by ":"
- Count the line number from 1
- Make sure to include the question number in your response
- Do not include the questions in your answer.
- Example:
    1. simple.func:1
    2. simple.examplefunc:2
    3. func:5

**Example {language} Code**:
```main.{language_extension}
{example code snippet}
```

**Example Questions**:
1. What are the fully qualified function calls at line 7 in file 'main.py'?
2. What are the fully qualified function calls at line 8 in file 'main.py'?

**Example Answers**:
1. main.simple_function:1
2. main.another_function:4

**{language} Code Provided**:

{code}

**Questions**:
{questions}

**Answers**:
{answers}
\end{lstlisting}
\end{minipage}
\section{Example Responses}
\label{sec:ex_responses}

\subsection{Type inference Output of mistral-large-it-2407-123b for test case args/assigned\_call}
\label{sec:response_mistral_types}
\begin{minipage}[b]{.5\textwidth}
	\captionsetup[lstlisting]{singlelinecheck=off,justification=raggedright}
	\begin{lstlisting}[language=Python, caption=Source code, style=pstyle, label=lst:types_ex_1_src]
def param_func():
    return "Hello!"


def func(a):
    return a()


b = param_func
c = func(b)
\end{lstlisting}

\end{minipage}%
\begin{minipage}[b]{.5\textwidth}
\captionsetup[lstlisting]{singlelinecheck=off,justification=raggedright}
	\begin{lstlisting}[language=json, caption=Raw LLM response, style=pstyle, label=lst:types_ex_1_res]
1. str
2. str
3. function
4. function
5. str
\end{lstlisting}

\end{minipage}%

\begin{minipage}[b]{.5\textwidth}
\captionsetup[lstlisting]{justification=raggedright}
\begin{lstlisting}[caption=Ground truth, language=json, label=lst:types_ex_1_gt]
[{
    "file": "main.py",
    "line_number": 1,
    "col_offset": 5,
    "function": "param_func",
    "type": ["str"]
},
{
    "file": "main.py",
    "line_number": 5,
    "col_offset": 5,
    "function": "func",
    "type": ["str"]
},
{
    "file": "main.py",
    "line_number": 5,
    "col_offset": 10,
    "parameter": "a",
    "function": "func",
    "type": ["callable"]
},
{
    "file": "main.py",
    "line_number": 9,
    "col_offset": 1,
    "variable": "b",
    "type": ["callable"]
},
{
    "file": "main.py",
    "line_number": 10,
    "col_offset": 1,
    "variable": "c",
    "type": ["str"]
}]
\end{lstlisting}
\end{minipage}
\begin{minipage}[b]{.5\textwidth}
	\captionsetup[lstlisting]{singlelinecheck=off,justification=raggedright}
\begin{lstlisting}[caption=Parsed JSON response, language=json, label=lst:types_ex_1_out]
[{
    "file": "main.py",
    "line_number": 1,
    "col_offset": 5,
    "function": "param_func",
    "type": ["str"]
},
{
    "file": "main.py",
    "line_number": 5,
    "col_offset": 5,
    "function": "func",
    "type": ["str"]
},
{
    "file": "main.py",
    "line_number": 5,
    "col_offset": 10,
    "parameter": "a",
    "function": "func",
    "type": ["callable"]
},
{
    "file": "main.py",
    "line_number": 9,
    "col_offset": 1,
    "variable": "b",
    "type": ["callable"]
},
{
    "file": "main.py",
    "line_number": 10,
    "col_offset": 1,
    "variable": "c",
    "type": ["str"]
}]
\end{lstlisting}
\end{minipage}

\subsection{Callgraph Output of mistral-large-it-2407-123b for test case args/assign\_return}
\label{sec:response_mistral}
\begin{minipage}[b]{.55\textwidth}
	\captionsetup[lstlisting]{singlelinecheck=off,justification=raggedright}
	\begin{lstlisting}[language=Python, caption=Source code, style=pstyle, label=lst:becnchmark_ex_1_src]
def return_func(args1):
    return args1


def func1():
    pass


def func2():
    pass

return_func(func1)
return_func(func2)()
\end{lstlisting}

\end{minipage}%
\begin{minipage}[b]{.45\textwidth}
\captionsetup[lstlisting]{justification=raggedright}
\begin{lstlisting}[caption=Ground truth, language=json, label=lst:res_1_gt]
{
  "main": [
    "main.return_func",
    "main.func2"
  ],
  "main.func1": [],
  "main.func2": [],
  "main.return_func": []
}
\end{lstlisting}
\end{minipage}

\begin{minipage}[b]{.55\textwidth}
\captionsetup[lstlisting]{singlelinecheck=off,justification=raggedright}
	\begin{lstlisting}[language=json, caption=Raw LLM response, style=pstyle, label=lst:becnchmark_ex_2_src]
1. main.return_func, main.func2
2.
3.
4.
\end{lstlisting}

\end{minipage}%
\begin{minipage}[b]{.45\textwidth}
	\captionsetup[lstlisting]{singlelinecheck=off,justification=raggedright}
\begin{lstlisting}[caption=Parsed JSON response, language=json, label=lst:res_1_out]
{
  "main": [
      "main.return_func",
      "main.func2"
  ],
  "main.func1": [],
  "main.func2": [],
  "main.return_func": []
}
\end{lstlisting}
\end{minipage}

\subsection{Callgraph Output of phi3-mini-it-3.8b for test case args/assign\_return}
\label{sec:response_phi}
\begin{minipage}[b]{.5\textwidth}
	\captionsetup[lstlisting]{singlelinecheck=off,justification=raggedright}
	\begin{lstlisting}[language=Python, caption=Source code, style=pstyle, label=lst:becnchmark_ex_3_src]
def return_func(args1):
    return args1


def func1():
    pass


def func2():
    pass

return_func(func1)
return_func(func2)()
\end{lstlisting}

\end{minipage}%
\begin{minipage}[b]{.5\textwidth}
\captionsetup[lstlisting]{justification=raggedright}
\begin{lstlisting}[caption=Ground truth, language=json, label=lst:res_1_gt]
{
  "main": [
    "main.return_func",
    "main.func2"
  ],
  "main.func1": [],
  "main.func2": [],
  "main.return_func": []
}
\end{lstlisting}
\end{minipage}

\begin{minipage}[b]{.5\textwidth}
\captionsetup[lstlisting]{singlelinecheck=off,justification=raggedright}
	\begin{lstlisting}[language=json, caption=Raw LLM response, style=pstyle, label=lst:becnchmark_ex_2_res]
1. func1, func2
2. func1
3. func2
4. return_func
\end{lstlisting}

\end{minipage}%
\begin{minipage}[b]{.5\textwidth}
	\captionsetup[lstlisting]{singlelinecheck=off,justification=raggedright}
\begin{lstlisting}[caption=Parsed JSON response, language=json, label=lst:res_1_out]
{
    "main": [
        "func1",
        "func2"
    ],
    "main.func1": [
        "func1"
    ],
    "main.func2": [
        "func2"
    ],
    "main.return_func": [
        "return_func"
    ]
}
\end{lstlisting}
\end{minipage}

\section*{Declarations}

\subsection*{Funding:}
Funding for this study was provided by the Ministry of Culture and Science of the State of North Rhine-Westphalia under the SAIL project with the grand no NW21-059D.

\subsection*{Ethical approval:}

Not applicable.

\subsection*{Informed consent:}

Not applicable.

\subsection*{Author Contributions}

\begin{itemize}
\item Ashwin Prasad Shivarpatna Venkatesh: First author, ideation, implementation, and execution of the whole idea.
\item Rose Sunil: Worked on the SWARM-JS part of the paper and its analysis.
\item Samkutty Sabu: Worked on the TypeEvalPy implementation and analysis of micro-benchmark results.
\item Amir M. Mir: Machine learning expert and worked on the analysis and gathering insights from observed data.
\item Sofia Reis: Static analysis expert and worked on the analysis and gathering insights from observed data.
\item Eric Bodden: Static analysis expert and worked on the analysis and gathering insights from observed data.
\end{itemize}

\subsection*{Data Availability Statements}
Data to reproduce experiments in this study, along with the source code, are published on GitHub at:
https://github.com/secure-software-engineering/TypeEvalPy and https://github.com/secure-software-engineering/SWARM-CG.
The raw outputs and analysis data is published on Zenodo at: https://zenodo.org/records/15045642

\subsection*{Conflict of interest}
The authors declare that they have no conflict of interest

\subsection*{Clinical Trial Number}
‘Clinical trial number: not applicable.

\bibliographystyle{ACM-Reference-Format}
\bibliography{references}

\end{document}